\documentstyle[12pt, epsf]{article}
\begin{document}

\newtheorem{lm}{Lemma}
\newtheorem{thm}{Theorem}
\newtheorem{cl}{Corollary}
\newcommand{\Al}{{\cal A} ({\bf G}) }
\newcommand{\f}{\psi}
\newcommand{\fb}{\bar{\psi}}
\newcommand{\zb}{\bar{z}}
\newcommand{\lv}{\langle vac |}
\newcommand{\rv}{| vac \rangle}
\newcommand{\G}{{\bf G}}
\newcommand{\rF}{{\cal F}_{R}}
\newcommand{\lF}{{\cal F}_{L}}
\newcommand{\ru}{| U \rangle}
\newcommand{\rs}{| S \rangle}
\newcommand{\llu}{\langle U |}
\newcommand{\lls}{\langle S  |}
\newcommand{\Gm}{{\bf G} _{-}}
\newcommand{\Gp}{{\bf G} _{+}}
\newcommand{\Pp}{P _{+}}
\newcommand{\Pm}{P_{-}}
\newcommand{\zv}{\vec{z}}
\newcommand{\wv}{\vec{w}}
\newcommand{\Vb}{\bar{V}}
\newcommand{\gl}{gl (V,  \bar{V})}
\newcommand{\Gl}{GL (V,  \bar{V})}
\newcommand{\X}{X (t,  \zv \,)}
\newcommand{\Xb}{\bar{X} (t,  \zv \, )}
\newcommand{\dq}{\frac{\partial}{\partial t _{n \cdot \ppb }}}
\newcommand{\llun}{\langle U_{N} | }
\newcommand{\run}{ | U_{N} \rangle}
\newcommand{\Uh}{\hat{U} _{N}}
\newcommand{\dv}{\vec{\partial}} 
\newcommand{\gb}{{\bf g}}
\newcommand{\hb}{{\bf h}}
\newcommand{\rb}{{\bf r}}
\newcommand{\qb}{{\bf q}}
\newcommand{\ppb}{{\bf p}}
\newcommand{\kb}{{\bf k}}
\newcommand{\w}{Diff( {\bf C})}
\newcommand{\B}{{\bf B}}
\newcommand{\I}{{\bf I}}
\newcommand{\Q}{{\bf Q}}
%{\obeylines
%\hfill UCDPHY-PUB-97-3
%\hfill UCDMATH-PUB-97-1
%\hfill February, 1997}
%\vskip 1.5 cm
\title{On the Structure of Correlation Functions \\
in the Normal Matrix Model }
\author{Ling-Lie  Chau$^{a}$ and Oleg  Zaboronsky$^{b}$ \\
$^{a}$Department of Physics,  University of California at Davis,  \\
Davis,  CA 95616, chau@physics.ucdavis.edu \\
$^{b}$Department of Mathematics,  University of California at Davis, \\
Davis,  CA 95616, zaboron@math.ucdavis.edu  }

\maketitle
\begin{abstract}
We study the structure of the normal matrix model (NMM).
We show that all correlation functions of the model 
with axially symmetric potentials can be
expressed in terms of holomorphic functions of one variable.
This observation is used to demonstrate the exact solvability
of the model.
The two-point correlation function is calculated in the  scaling limit
 by solving the  BBGKY\footnote{ Born, Bogoliubov, Green, Kirkwood,
 and Yvone, \cite{Mohling}.} chain of equations. The answer is shown to 
be
universal (i.e. potential independent up to a change of the scale). 
We then develop a two-dimensional free fermion formalism   and
construct a family of completely
integrable hierarchies (which we call
the extended-$KP(N)$ hierarchies)
of non-linear  differential equations.
The well-known KP hierarchy is a 
lower-dimensional reduction  of this family.  
The extended-$KP(1)$ hierarchy
contains the (2+1)-dimensional Burgers equations.
The partition function of the $(N\times N)$ NMM  
is the $\tau$ function of
the extended-$KP(N)$ hierarchy  
invariant with respect to a subalgebra of an algebra
of all infinitesimal diffeomorphisms of the plane. 
\end{abstract}
\vfill

\section{Introduction }
Normal Matrix Model (NMM) was introduced in \cite{Chau 1} and its 
connection
to
the Quantum Hall Effect was indicated.  In \cite{Chau 2} this connection 
was
given a precise form by observing that the partition function of NMM
coincides with zero-temperature partition function of two-dimensional
 electrons in the strong magnetic field. 

In the present paper we study the mathematical structure of the NMM.
The distinct feature of the NMM, in contrast to other matrix models
(Hermitian, Unitary, etc.),
is its relation to two-plus-one-dimensional [(2+1)-d] physical systems,
rather then the one-plus-one
-dimensional [(1+1)-d] ones. For example, we will exploit the equivalence of 
NMM
and the system of
(2+1)-d Coulomb particles to compute  correlation functions  of the NMM 
in
the scaling limit.
We will also show that the partition function of NMM is a
$\tau$-function of an integrable
hierarchy containing (2+1)-d Burgers equations.  

The paper is organized as follows. To make
the present exposition self-contained we
reproduce in Section 2
the results of  
\cite{Chau 2}, devoted to the definition of the model,
the derivation of the eigenvalue formula
for its partition function and the integrability of the model.
To emphasize the (2+1)-d nature of the NMM 
we notice following \cite{Chau 2}
that the partition function of the
model can be interpreted as
a classical partition function of Coulomb particles in a plane. In
contrast, the partition function of the 
Hermitian matrix model (HMM)  coincides with the classical
partition function of 
Coulomb particles constrained to a string in a plane.  
We show that the NMM
partition function
can be written in a determinant form.  
This enables us to conclude that the
partition function of the
model is a $\tau$-function of Toda lattice with respect to
holomorphic-antiholomorphic variations of
the potential. 

In Section 3 we study the NMM with axially-symmetric potential. First we
derive a 
 determinant representation for the correlation functions of the model.
 We notice that the orthogonal polynomials associated with the 
integration
measure
 are just powers of the complex variable $z$.  This simplifies the 
analysis of the correlation functions considerably. 
We compute the correlation functions for the case when the matrix model
potential is 
a monomial,  $V(M,  M ^{\dag}) =  (M \cdot M^{\dagger} )^{k}$. The 
answer is
expressed through
degenerate hypergeometric functions.  Thus NMM provides us with an 
example of
a matrix model
which is exactly (and explicitly) solvable beyond the Gaussian case even
before taking the continuum
limit.

The two-point function of Gaussian NMM decreases exponentially at
infinity,  which should
be compared to the power decay of the two-point function of HMM.  While 
the
latter can be obtained
from quantum-mechanical computation in the system of free $(1+1)$-d 
identical
fermions in the
external potential (see \cite{Mehta} for details), the former follows 
from
the analogous
computation in the system of free $(2+1)$-d identical fermions placed in 
the
external magnetic
field.

In general we find that all correlation functions can be expressed in 
terms
of a holomorphic
function in one variable.  This observation 
permits us to close a BBGKY 
chain of  equations for the
correlation functions and obtain a 
closed integro-differential equation for the
two-point function.  The existence of a single equation which determines
the two-point function and thus, through determinant representation, all
correlation functions, makes the NMM with axially symmetric potential
exactly solvable.

In Section 4 we solve the obtained equation
 in the continuum $\,N \rightarrow \infty$
limit.  The answer for the two-point function
is universal,  i.e.  potential-independent up to a change of  
scale. 

In Section 5, we develop a $(2+1)$-d free fermion formalism and
construct the free fermion representation of 
the partition function of NMM.
This fact has its analogy in the theory of HMM, 
in which the partition function admits
$(1+1)$-d free fermion representation. 

In Section 6,  we use the formalism developed in Section 5 to construct a
family of completely integrable
hierarchies (labeled by an integer $N$) of non-linear differential
equations. We show that the
$N=1$ hierarchy contains the
$(2+1)$-d Burgers equations (multidimensional
generalizations of the original Burgers equation
\cite{Burgers}, were considered 
firstin \cite{Gurb1}, see \cite{Gurb} for a review).
We thus call it the Burgers hierarchy. We give
explicit solutions to it.
These solutions generalize
the Hopf-Cole solutions
to the (2+1)-d Burgers equations,
\cite{Hopf} and \cite{Cole}.

The family constructed
provides  a multidimensional extension of KP
hierarchy which can be explained as follows.
It is well known that the  KP 
hierarchy
can be
formulated in terms of the pseudodifferential operator
$W(t,\partial )=1+w_1(t){\partial}^{-1}+w_2(t){\partial}^{-2}
 + \cdots$ (\cite{Sato} and
\cite{Sato1},
see \cite{JM} for a review). 
It is also known  
that KP hierarchy admits a reduction specified
by the condition that 
$W(t,\partial )\partial^N$ is a $ differential$ operator,
see \cite{Mulase}, \cite{Ohta} .We call 
it the 
$KP(N)$ hierarchy. The set of equations
of KP hierarchy coincides with that of 
$KP(N)$ hierarchy with $N \rightarrow \infty$.
We show that the $N$-th representative of
our family of hierarchies
of non-linear differential equations is a
multidimensional extension of $KP(N)$ hierarchy. Thus we call it the
extended-$KP(N)$
hierarchy.  In the simplest $N=1$ case 
the extended-$KP(1)$, or equivalently,
the Burgers hierarchy contains the
$(2+1)$-d Burgers equations, while the $KP(1)$
hierarchy contains the
$(1+1)$-d Burgers equation.

We then classify all formal solutions to the extended-$KP(N)$
hierarchy. We find the one-to-one correspondence between the
set of formal solutions and an open subset of an infinite-dimensional
Grassmann manifold $Gr (\infty , N)$. This subset consists of
all $N$-dimensional subspaces of 
an infinite-dimensional complex linear space having
a non-degenerate projection onto a fixed $N$-dimensional
subspace. 

Next we  show that the partition function of the
$N\times N$ NMM is a $\tau$-function of the
extended-$KP(N)$ hierarchy.
This is achieved by using  the bosonization
formula  which can regarded
as  a $(2+1)$-d generalization of $(1+1)$-d bosonization formulae. 

Finally  in Section 7 we discuss the Ward identities for the NMM.  It is
known that
matrix model solutions to the KP hierarchy (like HMM) satisfy Virasoro
constraints, i.e.  they are
annihilated by an infinite set of differential operators spanning a
subalgebra of Virasoro
algebra.  This subalgebra is isomorphic to
an algebra of infinitesimal {\em
holomorphic} polynomial
diffeomorphisms of the complex plane.  We show that the partition 
function of
NMM is a special solution to
the extended-$KP(N)$ hierarchy annihilated by
the set of differential operators generating an algebra 
of {\em all} infinitesimal polynomial
diffeomorphisms of the plane. 
The results of this Section can be used in the further
analysis of the continuum limit of the NMM.

The proofs of lemmas and theorems presented
in the paper
are placed between ``$\diamondsuit$"
signs.

\section{The Eigenvalue and the Determinant Forms of
the 
Normal Matrix Model  }

In this Section we will give the 
definition of the normal matrix model 
and derive the eigenvalue and 
determinant formulae for the 
partition function.   We will be utilizing the
standard methods of the theory of matrix models (see \cite{Morosov} for
review).   In the end of this Section,
we will derive a connection between 
NMM  and the Toda lattice hierarchy.  

The partition function of NMM has the following form:

\begin{eqnarray}
Z_{N} =\int_{\{ \Gamma : [M,   M^{\dag} ]=0 \} } d\mu (\Gamma )
e^{-trV(M,   M^{\dag})},\label{eqn;one}   
\end{eqnarray}
where $\Gamma$ denotes the set of 
$N \times N$ normal matrices,     $d \mu (\Gamma)$ 
is a measure  on $\Gamma$ induced by flat metric on the space
of all $N \times N$ complex matrices and   $V(z,  
\bar{z})$
is a function on ${\bf C}$ such that (\ref{eqn;one}) exists.  

As usual,    integral (\ref{eqn;one}) can be reduced to the integral over
eigenvalues $\{ z_{i} \}_{i=1}^{N}$ and $\{ \bar{z} _{i} \}_{i=1}^{N}$ 
of matrices $M$ and   $M^{\dag}$ respectively. 
A  corresponding calculation
can be easily
performed given the explicit expression for $d\mu (\Gamma )$
in 
the appropriate local coordinates on $\Gamma$.  So let us sketch the 
calculation
of  $d\mu (\Gamma )$.  

First we notice,    that any normal matrix  $M$ can be presented in the 
form
$M=UDU^{\dag}$,    where the $D$ is diagonal matrix,    $U \in U(N)$ is
a diagonalizing matrix.   This
decomposition
is unique up to multiplying $U$ by a diagonal unitary matrix from the 
right. 
The corresponding
equivalence class of $U$ together with $D$ defines a convenient 
coordinate
system on $\Gamma$.   Using these coordinates one can calculate the
Riemannian
metric
induced on $\Gamma$ by flat metric  on the space of all matrices:
\begin{eqnarray}
\parallel \delta M \parallel ^2 \equiv  tr \delta M \delta M ^{\dag} 
=tr (\delta D \cdot \delta D ^{\dag}) +
2tr(\delta u \cdot D \cdot \delta u \cdot D^{\dag}-
\delta u \cdot \delta u \cdot D \cdot D^{\dag}), \label{eqn;onea}  
\end{eqnarray}
where $\delta u = U^{\dag} \delta U$ is an invariant (Haar) length
element on $U(N)$. Due to the $U(N)$-invariance $\delta u$ 
is well-defined in terms of 
our coordinates on $\Gamma$.   Expressing  
$\parallel \delta M \parallel ^2$ through eigenvalues and matrix
elements $(\delta u)_{ij} \equiv   (U^{\dag} \delta U)_{ij}$ we get
\begin{eqnarray}
\parallel \delta M \parallel ^2 =\sum_{i=1}^{N} \delta z_{i} \delta 
\bar{z}
_{i}
-\sum_{i,   j=1}^{N} \delta u_{ij} \delta u_{ji} | z_{i}-z_{j} | ^2
.\label{eqn;two}
\end{eqnarray}

On the other hand,    $\parallel M \parallel ^2 =G_{ab} l^{a} l^{b}$,   
where
$\{ l^{a} \}$ is a cumulative notation for the local coordinates on 
$\Gamma,   G_{ab}$ is an induced metric on $\Gamma$.   Then
$\mu (\Gamma) =det(G)^{1/2} \prod_{a} dl^{a}$ 
(see e.g. \cite{Dubrovin}).   Combining these two
formulae with (\ref{eqn;two}) we obtain
\begin{eqnarray}
d\mu (\Gamma ) =dU \prod_{i=1}^{N} dz_{i} d \bar{z} _{i}
| \Delta (z) | ^2,\label{eqn;three}   
\end{eqnarray}
where $dU =\prod_{i \neq j} du_{ij} du_{ji}$ is Haar measure
on $U(N)$ and $\Delta (z) \equiv det [z_{i} ^{j-1}] _{1\leq i,j\leq N}
= \prod_{i > j} (z_{i} -z_{j})$ is the Van der Monde
determinant.  

Finally,    substituting (\ref{eqn;three}) into (\ref{eqn;one}) and
integrating over the unitary group
we arrive at the eigenvalue expression for $Z_{N}$:
\begin{eqnarray}
Z_{N} = c(N) \int_{{\bf C}^{N}}
( \prod_{i=1}^{N} dz_{i} d\bar{z} _{i}
e^{-V(z_{i},   \bar{z_{i}})})
| \Delta (z) | ^2,\label{eqn;four}   
\end{eqnarray}
where $c(N)$ is the volume
of the unitary group, a constant factor independent from $V$.
From now on we replace it by unity.  

Next we will deduce a determinant formula for the integral
(\ref{eqn;four}) and relate the result to the solutions to Toda lattice
hierarchy of differential equations.  

The reformulation of (\ref{eqn;four}) in determinant
form is standard and is based 
on the following formula:
\begin{eqnarray}
det [M_{ik}] \cdot det [M_{jk}] =\sum_{\sigma \in S^N} 
det [M_{i\sigma (j)} M_{j \sigma (j)} ] , \label{eqn;foura}  
\end{eqnarray}
where $M$ is any $N \times N$ matrix,    $\sigma$ is
an element of symmetric group $S^N$.   Applying 
(\ref{eqn;foura})
to the product of Van der Monde determinants
$\Delta (z) \Delta (\bar {z})$ in (\ref{eqn;four}) we arrive at the desired
form of the partition function NMM:
\begin{eqnarray}
Z_{N} &=& N! \cdot det[Z_{ij}]_{1\leq i, j\leq N}, \label{eqn;five}   
\mbox{ where } \\
Z_{ij} &\equiv & \int_{{\bf C}} dzd\bar{z} e^{-V(z,   \bar{z})} z^{i-1} 
\bar{z}
^{j-1} . \label{eqn;six} 
\end{eqnarray}

Let us consider the potential $V(z,   \bar{z})$ in the form
\begin{eqnarray}
V_{t}(z,   \bar{z})=U(z,   \bar{z})-\sum_{k > 0}(t_{k} z^k+\bar{t}_{k} 
\bar{z}^k)  \label{eqn;seven}
\end{eqnarray}
with $t_{k}=0=\bar{t}_{k}$ for $k \gg  1$.   We are interested in 
the behavior of $Z_{N} [V_{t}]$ with respect to variations of $t$'s and   
$\bar{t}$'s.   Note that $Z_{N} [V_{t}]$ can be considered as a 
generating
function of correlators  in NMM with
the potential $U(z,   \bar{z})$: 
\begin{eqnarray*}
 & & <(trM^i_{1})^{j_{1}}
\cdot (trM^{\dag i_{2}} )^{j_{2}} \cdots >_{NMM}  \\
 & & \equiv 
\frac{1}{Z_{N}}  \int_{{\bf C}^{N}}
\prod_{i=1}^{N} (dz_{i} d\bar{z} _{i}
e^{V(z_{i},   \bar{z_{i}})})
| \Delta (z) | ^2
\biggl( (trM^{i_{1}})^{j_{1}}
\cdot (trM^{\dag i_{2}} )^{j_{2}} \cdots \biggr) \\
 & & = \frac{1}{Z_{N}}(\frac{\partial}{\partial t_{i_{1}}})^{j_{1}} \cdot 
(\frac{\partial}{\partial \bar{t}_{i_{2}}})^{j_{2}} \cdots
Z_{N} [V_{t}] ~ \mid _{t, \bar{t}=0}.
\end{eqnarray*}
\noindent
From (\ref{eqn;six}) with potential given by 
(\ref{eqn;seven}) we can easily deduce that
\begin{eqnarray}
\frac{\partial Z_{ij}}{\partial t_{k}}=Z_{(i+k),j},   
\frac{\partial Z_{ij}}{\partial \bar{t}_{k}}=Z_{i,(j+k)}, ~~i,j\geq
1.\label{eqn;eight}
\end{eqnarray}
\noindent
Equation (\ref{eqn;five}) together with 
(\ref{eqn;eight}) mean that
$Z_{N}[V_{t}]$ is
an $N^{\rm th}$ $\tau$-function
 of  Toda lattice (with respect to complex variables
$t$'s and $\bar{t}$'s, see \cite{Leznov} and \cite{Ueno}  for
details). Note that the real version of
the integral (\ref{eqn;four}) together with 
potential (\ref{eqn;seven}) was studied in \cite{Morosov},
where it was referred to
as a ``scalar product model".  

In Section 6 we will
discuss a relation between NMM with an arbitrary potential
and integrable systems which includes the one described
above as a particular case.  

\section{Correlation Functions of the Normal Matrix Model}

In this Section we will analyze the structure of correlation
functions of NMM with arbitrary axially symmetric polynomial potentials.   
We will see in this case that the two-point correlation
function can be expressed in terms of the 
(analytical continuation of) one-point
function, thus leading to the exact solvability of the model.   

It follows from (\ref{eqn;four}) that the distribution function
of the eigenvalues in NMM with an axially symmetric potential
is 
\begin{eqnarray}
P_{N} (z, \bar{z} )=\frac{1}{Z_{N} } | \Delta (z) | ^2
 e^{\left(- \sum _{i=1}^{N} V(| z_{i} | ^2 
)\right)},\label{eqn;nine}   
\end{eqnarray}
where $Z_{N}$ is the partition function of 
NMM.   
The $n$-point correlation functions
(often called reduced distribution 
functions in statistical physics) are defined as follows:
\begin{eqnarray}
R_{N}^{(n)} (z_{1},   \cdots ,   z_{n} ) \equiv \frac{N !}{(N-n) !}
\int_{{ \bf C} ^{N-n}}
 \prod_{i=n+1}^{N} dz_{i}d \bar{z} _{i}
 P_{N} (z,   \bar{z} ), \label{eqn;ten}
\end{eqnarray}
where the combinatorial prefactor accounts for the symmetry of the 
integrand.
Note that
\begin{eqnarray}
R_{N}^{(1)} (z) = \langle \delta (z-z_{1} ) \delta (\bar{z} -
\bar{z}_{1} ) \rangle _{NMM}\label{eqn;tena}
\end{eqnarray}
and coincides therefore with the level density
of the NMM.  

Correlation functions (\ref{eqn;ten})
can be presented in the determinant
form. To prove it we note that 
the expression
(\ref{eqn;nine})
 for the distribution function can
be rewritten as
\begin{eqnarray}
P_{N}  (z,   \bar{z} )
=\frac{1}{N ~!} det  [K(\zb _{i},   z_{j})]_{1 \leq i,   j \leq N},
\label{eqn;eleven}
\end{eqnarray}
where
\begin{eqnarray}
K_{N} (\zb _{i},   z_{j}) \equiv
\sum_{k=0}^{N-1} \overline{ \phi} _{k} (\zb _{i})
\phi _{k} (z_{j}), \label{eqn;twelve}  
\end{eqnarray}
and
\begin{eqnarray}
\phi _{m} (z) \equiv \sqrt {c_m} z^{m} 
e^{- \frac{1}{2} V( | z | ^2)} , 
\label{eqn;thirteen}
\end{eqnarray}
where $c_{m} >0$.
The functions $\phi _{m} (z)$'s introduced above
are the orthogonal functions of
the problem
normalized by the following condition:
\begin{eqnarray}
\int_{ {\bf C} } dz d \bar{z} 
\overline{\phi} _{m} (\zb) \phi _{m'} (z)
=\delta _{m,m'},\label{eqn;thirteena}
\end{eqnarray}
which yields
\begin{eqnarray}
\pi c_{m} \int_{0}^{\infty} dx ~ \left[x^{m} e^{ -V(x)}\right] =1
.\label{eqn;sixteen}
\end{eqnarray}
Apparently, the orthogonal polynomials associated to the NMM with
axially symmetric potential are just monomials in $z$. This
explains the certain simplicity of the model.   

It is easy to check that the matrix $K_{N} (\zb _{i},   z_{j})$ is
hermitian and it can be considered
as the kernel of projection operator,    i.e.   the relation
\begin{eqnarray}
\int_{ {\bf C} }
dv d \bar{v} K_{N} (\bar{u},   v) K_{N} (\bar{v},   w) 
= K_{N} (\bar{u},  w) 
\label{eqn;thirteenb}
\end{eqnarray}
 is satisfied.  Therefore one can 
make use of the well-known result from the theory of
matrix models (see \cite{Mehta},    page 89,    theorem 5.2.1)
to obtain the following representation of the $n$-point
functions defined in (\ref{eqn;ten}):  
\begin{eqnarray}
R_{N}^{(n)} (z_{1},   \cdots ,   z_{n} ) 
=det [K_{N} (\zb _{i},   z_{j} )]_{1 \leq i,   j \leq n} 
.\label{eqn;fourteen}
\end{eqnarray}
\noindent
It is convenient to rewrite the expression
for $K_{N} (\zb _{i} ,   z_{j} )$ by substituting (\ref{eqn;thirteen}) into
(\ref{eqn;twelve}):
\begin{eqnarray}
K_{N} (\zb_{i},   z_{j}) &=&
 \sum_{m=0}^{N-1} c_{m}\cdot (\bar{z} _{i} \cdot z_{j} )^m 
e^{\frac{1}{2} \left(-V( | z_{i} | ^2) -V( | z_{j} | 
^2)\right) }
\nonumber \\
&\equiv& k_{N} (\overline{z_{i}} \cdot
z_{j} ) e^{  \frac{1}{2} \left(-V(| z_{i} | )^2
- V(| z_{j} | ) ^2 \right) }.   
\label{eqn;fifteen}
\end{eqnarray}
Therefore, up to a factor which explicitly depends on the 
potential,   
$K_{N} (\zb ,   w)$ is completely determined by the function of one complex
variable $k_{N} (\bar{z} \cdot w)$.  

To illustrate  the formulae obtained above
let us consider the NMM 
with the Gaussian potential, $V(| z | ^2) =  | z 
| ^2$, in the limit $N \rightarrow \infty$.   A simple calculation 
gives the following answer for the kernel:
\begin{eqnarray}
K (\zb ,   w) = \frac{1}{\pi} e^{\left(\bar{z} w
-\frac{1}{2} | z | ^2 -\frac{1}{2} | w | ^2\right)
}.\label{eqn;seventeen}
\end{eqnarray}
Substituting (\ref{eqn;seventeen}) into (\ref{eqn;fourteen})
we conclude that in the limit $N \rightarrow \infty$
the one-point correlation function
is constant and
 equals to $\frac{1}{\pi}$,    whereas the 
2-point correlation function in the same limit is 
\begin{eqnarray}
R^{(2)}(z,   w)=\frac{1}{\pi ^{2}} [1-e^{ (- | z -w | ^2 )}
].\label{eqn;eighteen}
\end{eqnarray}
It follows from the last two results  that  the connected part of the 
2-point function is
equal to $\frac{1}{\pi ^2} exp[- | z-w | ^2]$ 
and decays exponentially at infinity.  
Let us note that the Gaussian
NMM is equivalent to the Gaussian Complex matrix model
(see \cite{Mehta},    Chapter 15). However this
is no longer true for a non-Gaussian matrix model potential.   
  
Computability of NMM
goes beyond the Gaussian case.  For example it is
 possible to obtain a closed expression for $K (\zb ,   w)$
when the NMM potential is a monomial in $| z | ^2:
V(z, \bar{z})= | z | ^{2k}$,    where $k$ is a positive integer.   
Corresponding
 integral
in (\ref{eqn;sixteen}) is easily calculated to give the following 
expression
for $c_{m}$:
\begin{eqnarray}
c_{m}=\frac{k}{\pi} \frac{1}{\Gamma ( \frac{m+1}{k} ) }. 
\label{eqn;nineteen}
\end{eqnarray}
Substituting (\ref{eqn;nineteen}) into (\ref{eqn;fifteen}) one obtains 
the
following
answer for the kernel at $N = \infty$:
\begin{eqnarray}
K (\zb ,   w) = \sum_{p=0}^{k-1}
 \frac{ (\bar{z} w)^p }{\Gamma ( \frac{p+1}{k})}
F_{1,   1} \left(1;\frac{p+1}{k} ; (\bar{z} w)^k \right)
e^{(- \frac{1}{2} | z |
^{2k} -
\frac{1}{2} | w | ^{2k}  )},\label{eqn;twenty}   
\end{eqnarray}
where $F_{1,   1} (a;b;z)$ is the degenerate hypergeometric function,   
the solution to the Kummer equation which is
analytic  at $z =0$:
\begin{eqnarray}
z\frac{d^2 \omega}{d z^2} + (b-z) \frac{d \omega}{d z} -a \omega
=0.\label{eqn;twentya}
\end{eqnarray}
For $k=1$, (\ref{eqn;twenty}) reduces to (\ref{eqn;seventeen}).  
It is well known (see e. g. \cite{Gasper}) 
that the $F_{1,   1} (a,   b,   z)$ has a Taylor
expansion in $z$ with an infinite radius of convergence and defines
therefore
an entire function on the complex plane. We use this remark below to
analyze the correlation functions of NMM in the case of an arbitrary
polynomial potential $V$.   

Suppose $V(x)=\sum_{n=1}^{d} t_{n} x^{n}$ with $t_{d} >0$.   
Then there are positive constants $a_{1}$ and  $ a_{2}$ such that
\begin{eqnarray}
V(x) \leq a_{1} + a_{2} x^{d}~,   ~ x \geq 0 . \label{eqn;twentyb}
\end{eqnarray}
This estimate can be used to prove that for any complex
number $u$ and a positive number $R$ such that $| u | \leq R$,
\begin{eqnarray}
\bigg| k_{N} (u)_{~{\rm for}~ 
V(x)=\sum_{n=1}^{d} t_{n} x^{n }} \bigg|~   \leq e^{-a_1}~ \bigg|~
k_{N} (R)_{~{\rm
for}~ 
V(x) = a_{2} x^{d}}\bigg|.\label{eqn;twentyone}
\end{eqnarray}
But the r.h.s. of (\ref{eqn;twentyone}) is nothing but a partial sum of a
series
converging by the remark above for any $0 \leq R < \infty$.   
Therefore $ k (u) \equiv \lim_{N \rightarrow \infty} k_{N}$
exists for any $u$ such that $| u | < \infty$ and defines
an entire function on the complex plane.   But an entire function
is uniquely determined by its values on say positive
part of the real line,    which
provides us with the connection between 2- and 1-point
correlation functions.   To make
this connection explicit  we notice that by definition
\begin{eqnarray}
R^{(1)} (| z |^2) =k\left(| z |^2 \right) e^{-V\left( | z
|
^2\right)}.\label{eqn;twentytwo}
\end{eqnarray}
Denote by $R^{(1)} (u)$ an entire function of a complex variable $u$
which coincides with the 1-point function above for $u= | z |^2$.   
Such function exists as $V$ is a polynomial and $k (u)$ has been proved
to be an entire function.   Moreover it is unique,    therefore is 
determined
completely by the 1-point function.  We finally notice that the connected
part of the 2-point function can be expressed in turn in terms of 
$R^{(1)}
(u)$:
\begin{eqnarray}
R_{c}^{(2)} (z,   w) = \bigg| R^{(1)} (\bar{z} w) \bigg|^2 
e^{-\left[V( | z | ^2) + V(| w | ^2) -V(\bar{z} w) -V(z
\bar{w})\right]},
\label{eqn;twentythree}
\end{eqnarray}
which states the claimed property of the NMM.  It's worth
mentioning that there is 
the counterpart of (\ref{eqn;twentythree}) for any finite $N$ as well.

Relation (\ref{eqn;twentythree}) leads to an exact solvability of NMM.  
Let
us discuss this
point in some more details.  
It follows from (\ref{eqn;fourteen})
and (\ref{eqn;fifteen}) that all 
correlation functions of the model can be
expressed in terms of $R^{(1)} (u)$.  On the other side
they can be considered as a solution to the
certain chain of integro-differential
equations which is discussed below.  These two remarks
permit one to derive a closed equation for the function
$R^{(1)} (u)$ which is what we mean claiming the
exact solvability of NMM.  To obtain such an
equation we rewrite an expression for the distribution function
(\ref{eqn;nine}) in the following way:
\begin{eqnarray}
P_{N} (z,   \bar{z}) =\frac{1}{Z_{N} }
e^{\left(-\sum_{i=1}^{N} V(| z_{i} | ^2 ) + 2 \sum_{i < j}
ln | z_{i} - z_{j} | \right)}.\label{eqn;twentyfour}
\end{eqnarray}
(\ref{eqn;twentyfour}) can be identified with the distribution
function
of classical two-dimensional Coulomb gas in equilibrium.  
The analogous equivalence has been widely exploited
in the study of conventional matrix ensembles (see
\cite{Mehta} for review).     

Such an identification allows one to use the Liouville's Theorem of
classical mechanics to derive a BBGKY chain
of equations for the reduced distribution functions
(\ref{eqn;ten}) (see e.g.~\cite{Mohling}).  In particular we
have the following equations,   connecting
one- and two-point functions:
\begin{eqnarray}
\frac{\partial R_{N} ^{(1)} (|z|^2)}{\partial z} +
R_{N}^{(1)} (|z|^2) \frac{\partial V (z) }{\partial z}
= \int dw d\bar{w}~ R_{N}^{(2)} (z,   w) 
\frac{\partial}{\partial z}
ln | z-w | ,\label{eqn;twentyfive}    
\end{eqnarray}
and another one obtained from (\ref{eqn;twentyfive}) by means of complex
conjugation.  Substituting (\ref{eqn;twentythree}) into
(\ref {eqn;twentyfive}) we get a closed equation for $R^{(1)} (u)$:
\begin{eqnarray}
\frac{\partial R_{N} ^{(1)} (|z|^2)}{\partial z} +
R_{N}^{(1)} (|z|^2) \frac{\partial V (z) }{\partial z}
= R_{N}^{(1)} (| z| ^2 ) 
 \int dw d\bar{w}~ R_{N}^{(1)} (| w| ^2 ) 
\frac{\partial}{\partial z}
ln | z-w | & & \nonumber \\ 
-  \int dw d\bar{w}~\bigg| R^{(1)} (\bar{z} w) \bigg|^2 
e^{-\left[V( | z | ^2) + V(| w | ^2) -V(\bar{z} w) -V(z
\bar{w})\right]}
 \frac{\partial}{\partial z} ln | z-w | & & .
\label{eqn;twentyfiveaa}    
\end{eqnarray}
Equation (\ref{eqn;twentyfiveaa}), together with a reality condition
$\overline{R^{(1)} (u)} = R^{(1)} (\bar{u})$
and normalization condition $\int 
dz d
\bar{z}
R_{N}^{(1)} (|z|^2)=N,$ can be used to
determine the entire
function $R_{N}^{(1)} (u)$ and thus all correlation
functions of NMM.  
In what follows we will exploit  (\ref{eqn;twentyfiveaa}) to study the 
$N \rightarrow \infty$ scaling limit of NMM.  

\section{NMM in the Continuum Limit  }

The present Section is devoted to the NMM
in $N \rightarrow \infty$ scaling (or continuum) limit.    We
will find the corresponding limits of level density and 
the two-point correlation
function.   We will see that in the regions where the limiting
level density is non-zero and smooth (no points of phase
transition) the form of the two-point
function is universal, i.e.   independent
of the NMM potential up to a rescaling .
The scale itself is determined by the 
level density which happens
to be inversely proportional to the 
square of correlation length.

It is appropriate to mention that our search
for universal answers in the NMM
has been motivated by the 
study of universality in the hermitian matrix models,
see e. g.  \cite{BZ0}.

Consider the $N \times N$ normal matrix model 
with the potential $N \cdot V(| z | ^2)$,   
where $V(x)$ is a real polynomial with
positive coefficient
in front of the monomial of the highest degree.
We assume $V$ to be  ``convex," by
which we mean that the following inequality holds for
any $z, w \in {\bf C}$:
\begin{eqnarray}
 V( | z | ^2) + V(| w | ^2) -V(\bar{z} w) -V(z \bar{w}) \geq 0,
\label{conv}
\end{eqnarray}
and the equality in (\ref{conv}) is reached iff $z=w$. 
Setting $w=z+\epsilon$
and expanding the l.h.s. of (\ref{conv}) around $\epsilon =0$ we get
the infinitesimal version of  the convexity condition:
\begin{eqnarray}
\Delta V(| z | ^2 ) \geq 0,
\end{eqnarray}
where $\Delta \equiv \partial \bar{\partial}$ 
is a two-dimensional Laplace operator. 
It is easy to check that any monomial 
potential of positive degree satisfies (\ref{conv}),
therefore any polynomial of  positive 
degree with positive coefficients does.

We will
be interested in the asymptotic properties
of the model as $N \rightarrow \infty$.
We will consider two cases. First, we consider the limit $N \rightarrow
\infty$ when the range of  variation of
 $ | z |^2 $ is of order $ 1$, which corresponds to the limit of strong
potential and
large
separation of levels. Second, we will investigate the case of 
strong potential and small separation of levels, i. e.
the limit $N \rightarrow \infty$ when the range of variation
of $| z |^2$ is of order $\frac{1}{N}$.

For the first case we will seek the solution
to (\ref{eqn;twentyfiveaa}) in the form of the asymptotic
expansion in the inverse powers of $N$,
\begin{eqnarray}
R ^{(1)}_{N} (u) = N \sum_{n=0}^{\infty} R^{(1)}_{n} (u) N^{-n}.
\label{as}
\end{eqnarray} 
The limiting one-point function is defined to be
\begin{eqnarray}
R^{(1)} (u) = R^{(1)}_{0} (u) \equiv \lim_{N \rightarrow \infty}
\frac{R^{(1)}_{N} (u)}{N} . 
\label{lim}
\end{eqnarray}
Substituting (\ref{as}) into (\ref{eqn;twentyfiveaa})  and equating the 
terms of the
first order in $N$ we get the following equation for  
the limiting one-point function:
\begin{eqnarray*}
R^{(1)} (|z|^2) \biggl( \frac{\partial V (|z|^2) }{\partial z}
- \int dw d\bar{w} R^{(1)} ( |w|^2) \frac{\partial}{\partial z}
ln | z-w | \biggr) =0 ,    
\end{eqnarray*}
which implies that
\begin{eqnarray}
\frac{\partial V (|z|^2) }{\partial z}
= \int dw d\bar{w} R^{(1)} ( |w|^2) \frac{\partial}{\partial z}
ln | z-w | ,
\label{R}    
\end{eqnarray}
if we assume that $ R^{(1)} (|z|^2) \neq 0$.
To derive (\ref{R}) from (\ref{eqn;twentyfiveaa})
we have used the convexity condition
and relation (\ref{eqn;twentythree}). Relation (\ref{eqn;twentythree})
was used  to show that in the limit considered
the connected
part of the two-point function is zero almost everywhere.  
Note that we derived (\ref{eqn;twentyfiveaa}) from the
Liouville theorem, so  its direct consequence, equation (\ref{R}),
 provides an exact description of the continuum limit of NMM.

It is interesting to note that equation (\ref{R}) can be interpreted
as an extremum
condition for the potential energy $U [\rho]$ of
the gas of charged particles of density $\rho (|z|^2)=R^{(1)} (|z|^2)$  
placed in the external potential:
\begin{eqnarray}
U [\rho ] = \int dz d \zb \rho V (|z|^2)
- \frac{1}{2} \int \int dz d \zb dw d \bar{w} 
\rho (|z|^2) \rho (|w|^2) ln | z- w | ^{2}.
\end{eqnarray}
Such an interpretation was a starting point in the derivation
of the level density in the hermitian matrix model
presented in \cite{Mehta}, however it required an
assumption that the two-point function
factorizes in $N \rightarrow \infty$ limit. In contrast,
in the NMM we can demonstrate the 
factorization of the two-point function 
almost everywhere by analyzing (\ref{eqn;twentythree}).  

Equation (\ref{R}) can be easily solved  by differentiating
both sides with respect to $\bar{z}$. Noticing that
$ln |z- w|$ is a Green function of two-dimensional
Laplacian, i.e. $\Delta ln |z-w| =  \pi \delta ^{(2)} (z-w)$,
we get
\begin{eqnarray}
R^{(1)} (|z|^2) = \frac{1}{\pi} \Delta V(|z|^2).
\label{density}
\end{eqnarray}  
So we conclude that the limiting level density (one-point
function)
of NMM  at the point $z$ is equal to $\frac{1}{\pi} \Delta
V(| z | ^{2})$,   given that the limiting
level density is not zero.
This answer has a transparent physical meaning: the
gas of charged particles which describes the continuous limit
of NMM tends
to screen completely the external potential.

Having described the spectrum locally one can derive a complete
picture assuming that the potential energy of the ``gas" of NMM
eigenvalues is minimal. In our case $V$ is convex
and the minimum energy assumption yields 
the following answer for the level density:
\begin{eqnarray}
R^{(1)} (|z|^2) &=& \frac{1}{\pi} \Delta V(|z|^2)
 \mbox{ if } | z | ^2 < r \label{eqn;twentyeighta} \\
R^{(1)} (|z|^2) &=& 0 \mbox{ if } | z | ^2 > r,   \label{eqn;twentynine}
\end{eqnarray}
where the end point $r$ of the spectrum is determined
from the normalization condition
\begin{eqnarray}
\int_{| z | ^2 <r } dz d \bar{z} R^{(1)} (|z|^2) 
=1.\label{eqn;twentyninea}
\end{eqnarray}

Consider now the case of small distances between
eigenvalues, for which the two-point correlations are
essential.   
Let $x$ be a real positive number. 
We assume that the $R^{(1)}(x) > 0$
and there is a closed neighborhood of 
$x$ in the complex plane where
the convergence of the r.h.s. 
of (\ref{lim}) is uniform (in physical terms,
$x$ is not a point of a phase transition 
in the $N \rightarrow \infty$ limit). 
We will study the limiting behavior of the two-point function
of NMM in the vicinity of such a point. Note that
the existence of points of uniform convergence
does not contradict the information about the structure
of correlation functions we have collected so far. For
example, the uniform convergence of the sequence of entire
functions $R_{N}^{(1)}$ in the closed neighborhood $U$
of $x$ implies that the limit is a holomorphic function
in this neighborhood. But according to (\ref{density})
$R^{(1)} (x)$ admits an analytic continuation
in some neighborhood of $x$ given that $R^{(1)} (x) \neq 0$
(recall that $V(x)$ is a polynomial function). Moreover,
a direct
verification shows that  
in the case of monomial potentials
$x$ is a point of uniform convergence
if  $x$ is not too large
and $R^{(1)} (x) \neq 0$. We conclude therefore
that the domain
of our consideration is not empty. However,
it does not consist of the whole plane: 
it is apparent that the endpoints of the spectrum 
defined by (\ref{eqn;twentyninea}) are
excluded from our considerations, as the sequence of continuous functions
cannot converge uniformly to a discontinuous one. Thus the
results concerning the two-point functions presented
below are valid only in the ``bulk" of the spectrum.

Consider now the following holomorphic change of
coordinates:
\begin{eqnarray}
z_{N} = \sqrt{x} \cdot e^{\zeta / \sqrt{N}},~
w_{N} = \sqrt{x} \cdot e^{\eta / \sqrt{N}},~
u_{N} \equiv \bar{z}_{N} \cdot w_{N}
.\label{eqn;twentyfivea}
\end{eqnarray}
The corresponding limit of the one-point 
function is $R^{(1)} (x, \bar{\zeta} +\eta ) \equiv
lim_{N \rightarrow \infty} \frac{R^{(1)}_{N} (u_{N})}{N}$.
It is a matter of simple considerations to show that
if $\lim_{N \rightarrow \infty} g _{N} (u) = g (u)$
 uniformly, $u \in U$, then 
$\lim_{N \rightarrow \infty} g _{N} 
(u \cdot f_{N}) = g (u)$ for any sequence $\{ f_{N} \}$
converging to $1$ and such that $u \cdot f_{N} \in U$ for any $N$.
Using this remark we see that
$R^{(1)} (x, \bar{\zeta} +\eta ) = R^{(1)}(x)$.
$\biggl[$ Let us also remark 
that the expression $R^{(1)} (x, \bar{\zeta} +\eta ) = R^{(1)}(x)$
solves the limit of the equation 
(\ref{eqn;twentyfivea}) corresponding
to small separations between levels.
One can consider this remark as yet
another consistency check of the assumption
about the existence of the point of uniform
convergence.$\biggr]$
Now we can substitute
(\ref{eqn;twentyfivea} into (\ref{eqn;twentythree})
and  take the limit $N \rightarrow
\infty$ of (\ref{eqn;twentythree}) to obtain
the following answer for the scaling limit
of the two-point function:
\begin{eqnarray}
R^{(2)}_{c} (x, \zeta  ,   \eta) = [R^{(1)} (x)]^2 
e^{-\pi \cdot R^{(1)}(x) | \zeta- \eta | ^2 },   \label{eqn;thirty}
\end{eqnarray}
where we used the result (\ref{density}) and
assumed that $R^{(1)} (x) \neq 0$.  
We conclude from (\ref{eqn;thirty}) that the scaling 
limit of the two-point function of NMM
exhibits universal behavior since the dependence 
on potential (which enters the answer
through $R^{(1)} (x)$ only) 
can be eliminated by changing
the scale and normalization.
Notice that the square of correlation length is
given by inverse level density.  
Equation (\ref{eqn;thirty}) can also be interpreted 
as a completely universal
relation between one- and two-point functions
of the NMM.  A similar phenomenon was
previously observed in \cite{BZ} for the hermitian
matrix model and is referred to as ``universality
of the second type".

\section{Free Fermion Representation of NMM }

The aim of the present Section is to rewrite the partition
function (\ref{eqn;one}) of NMM in the form of the free fermion 
correlator. 
This
will permit us to describe an arbitrary variation of NMM potential
by means of integrable system of differential equations and generalize
therefore
the results of Section 2.  
The technique we use
is a straightforward generalization of free fermion methods
developed in 
\cite{SJM} and reviewed in part in \cite{JM} 
and their application to the theory of
matrix models 
due to the ITEP group (see \cite{Morosov} for review).

Consider an abelian group ${\bf Z} \times {\bf Z}$ consisting
of pairs of integers.
The
corresponding group multiplication will be
denoted by ``$ + "$.
To simplify notations
we refer to this group as ${\bf G}$. 
An abelian group ${\bf Z}$ acts on ${\bf G}$
as follows:
\begin{eqnarray}
{\bf G}  \times {\bf Z} &\rightarrow& {\bf G},\nonumber \\
\biggl(\gb =(i, j), m\biggr) &\mapsto& \gb \cdot m  = (i \cdot m ,  j \cdot
m).\label{eqn;thirtya}
\end{eqnarray}
Alternatively, ${\bf G}$ can be presented in the following
form:
\begin{eqnarray}
{\bf G}= \oplus _{m \in {\bf Z}} {\bf G}_{m} = \Gm \oplus \Gp,
\label{eqn;thirtyone}
\end{eqnarray}
\noindent
where
\begin{eqnarray}
\Gm = \oplus _{m < 0} {\bf G}_{m} ~ ~ ; ~  ~ \Gp = \oplus _{m \geq 0}
{\bf G}_{m}, \label{eqn;thirtytwo}
\end{eqnarray}
and
\begin{eqnarray}
{\bf G}_{m} =\biggl\{ \gb \in {\bf G} \bigg| 
\gb =(l+m, -l) ~;~ l \in {\bf Z} \biggr\}~,~ m
\in {\bf Z}.
\label{eqn;thirtythree}
\end{eqnarray}
Note that ${\bf G}_{0}$ can be visualized 
as an antidiagonal of ${\bf G}={\bf Z} \times {\bf Z}$. 
It is also worth mentioning that
if $\gb \in {\bf G}_{m}~\mbox{and} ~ \hb  \in {\bf G}_{n}$, then $\gb 
+\hb \in {\bf
G}_{n+m}$. Therefore,
${\bf G}$ is
${\bf Z}$-graded and (\ref{eqn;thirtyone}) is a decomposition of ${\bf 
G}$
into a sum
of components
of a fixed degree. 
 
Let us consider an infinite-dimensional
 Clifford algebra $\Al$ over
complex numbers which corresponds to ${\bf G}$. We define it
 by means of
the following set of generators and relations:
\begin{eqnarray}
\Al =\biggl \langle 1,    \f _{\gb},    \fb _{\hb} ; \gb, \hb \in {\bf G} 
%|
\bigg|
\{ \f _{\gb } ,    \fb _{\hb } \} =\delta _{\gb , \hb } ,   
\{ \f _{\gb } ,    \f _{\hb } \} =0= \{ \fb _{\gb } ,   \fb _{\hb } \}
\biggr\rangle.\label{eqn;thirtyfour}
\end{eqnarray}
\noindent
Let $\rF$ be the right Fock space,
an irreducible left $\Al$-module
obtained by applying $\Al$ to the right vacuum vector $\rv$, which
is defined as follows:
\begin{eqnarray}
\f _{\gb } \rv =0, ~ \gb  \in \Gm ~;\quad ~ \fb _{\hb } \rv =0, ~ \hb  
\in
\Gp .\label{eqn;thirtyfive}
\end{eqnarray}
In the same fashion we introduce the  left Fock space  $\lF$ which
is generated by the right action of $\Al$ on the left vacuum vector
$\lv$ which is defined below:
\begin{eqnarray}
\lv  \fb  _{\gb } =0, ~ \gb  \in \Gm ~ ;~ \lv  \f_{\hb } =0, ~ \hb  \in 
\Gp
.\label{eqn;thirtysix}
\end{eqnarray}
There is a natural non-degenerate
pairing between left and right  Fock
spaces,
\begin{eqnarray*}
\biggl(\lv a ,    b \rv \biggr) \mapsto \lv a \cdot b \rv,
\end{eqnarray*}
where $a$ and
$b$ are elements of $ \Al$.
This pairing is normalized as follows:
\begin{eqnarray}
\lv 1\cdot 1 \rv =1 .\label{eqn;thirtyseven}
\end{eqnarray}

\noindent
From now on we will think of $\Al$ as an algebra of linear
transformations of $\rF$ or $\lF$ and refer to elements
of $\Al$ as operators. 

To each element $a \in \Al$ one can assign its
normal reordering with respect to the vacuum
$\rv$, an element of $\Al$,
 denoted as $:a:$.  It is obtained
from an element $a$ by permuting the generators in each term
of the sum constituting the element $a$ 
in such a way that the fermion generators annihilating the right
vacuum stand to the right  from the rest of
generators.  

A variety of vectors in the Fock space is provided by
means of the following construction. 
To each finite set $S \subset {\bf G}$ we assign the following vector in
$\rF$:
\begin{eqnarray}
\rs = \prod_{\gb \in S}  o _{\gb} \rv ,
\label{eqn;thirtynine}
\end{eqnarray}
where $ o  _{\gb} = \f _{\gb}$ if $\gb \in \Gp \bigcap S$
and $ o _{\gb} = \fb _{\gb}$ if $\gb \in \Gm \bigcap S$. 
Note that without fixing a linear
order in the set ${\bf Z} \times {\bf Z}$,    the element
(\ref{eqn;thirtynine}) is defined up to a sign only.  
The same construction can be applied
to obtain a vector in $\lF$ which we will
denote as $\lls$, 
\begin{eqnarray}
\lls = \lv \prod_{ \gb \in S} 
\bar{ o} _{\gb},
.\label{eqn;forty}
\end{eqnarray}
where $\bar{ o}  _{\gb} = \fb _{\gb}$ if $\gb \in \Gp \bigcap S$
and $\bar{ o} _{\gb} = \f _{\gb}$ if $\gb \in \Gm \bigcap S$. 
It is easy to see that any element of $\rF$ or $\lF$  can be
presented as a linear
combination of vectors (\ref{eqn;thirtynine}) or (\ref{eqn;forty}),
respectively.

Consider a semigroup ${\bf Q} \subset {\bf G}$ 
consisting of all pairs $\gb =(m,n)$,
where $m$ and $n$ are non-negative integers. 
We introduce the following element 
of the algebra $\Al$:
\begin{eqnarray}
H(t) =\sum_{\gb  \in {\bf Q}}  t_{\gb } J_{\gb } \equiv \sum_{\gb  \in 
{\bf
Q} }
t_{\gb }
(\sum_{\hb  \in {\bf G}} \f _{\hb } \fb _{\hb  +\gb }),   
\label{eqn;fortyone}
\end{eqnarray}
where 
\begin{eqnarray}
t_{\gb } =0 \mbox{ if } deg (\gb ) >> 0,  
\label{eqn;fortytwo}
\end{eqnarray}
and we always suppose that $t_{\bf{0} }=0$. 
 We will call $H(t)$ a Hamiltonian operator
or simply Hamiltonian.  
It is easy to check that the operators $J_{\gb }$'s introduced
in (\ref{eqn;fortyone}) commute,  i.e.  $[ J_{\gb } ,  J_{\hb }]
\equiv J_{\gb } \cdot J_{\hb } - J_{\hb } \cdot J_{\gb }=0$,
for
$\gb ,~ \hb  \in {\bf Q} \setminus \{ \bf{0} \}$.
It is also a matter of simple computation to verify
that $[J_{\gb} , \f_{\hb}] = \f _{\hb -\gb}$ and
$[J_{\gb} , \fb _{\hb}]= \fb _{\hb + \gb}$. 

Before we introduced the Hamiltonian (\ref{eqn;fortyone})
all our considerations had not been
different from the standard fermion construction as
$\Al$ is isomorphic to a standard Clifford
algebra the generators of which are labeled by elements
of ${\bf Z}$, the isomorphism is being
established by ordering the set ${\bf Z}
\times {\bf Z}$. The new development starts with
the introduction of Hamiltonian (\ref{eqn;fortyone}) since the presented
commutation
relations between $J_{\gb}$'s and fermion generators
depend on the group structure in ${\bf G}= {\bf Z} \times
{\bf Z}$, but $\bf{G}$ is not isomorphic to  ${\bf Z}$ as a group.      
 
It follows from the definition (\ref{eqn;fortyone})  that the Hamiltonian
operator annihilates vacuum, 
 \begin{eqnarray}
H(t) \rv =0.  \label{eqn;fortythree} 
\end{eqnarray}
\noindent  
Define a $t$-evolution of an element
$a \in \Al$ as the following
element of (the formal completion of) $\Al$:
\begin{eqnarray}
a(t)=e^{H(t)} a e^{-H(t)}.\label{eqn;fortyfour}
\end{eqnarray}
\noindent
Due to (\ref{eqn;fortythree}) the $t$-evolution preserves the normal ordering,
$: a(t) : =:a: (t)$. It is not difficult
to derive the following expressions
for the $t$-evolution of fermionic generators of $\Al$:
\begin{eqnarray}
\f _{\gb } (t) =\sum_{\hb  \in {\bf Q}} p _{\hb } (t) \f _{\gb -\hb },
\label{eqn;fortyfive}\\
\fb_{\gb } (t) =\sum_{\hb  \in {\bf Q}} p _{\hb } (-t) \fb _{\gb +\hb },
\label{eqn;fortysix}
\end{eqnarray}
where $\{ p_{\gb } \} _{\gb  \in {\bf Q}}$ are {\em generalized
Schur polynomials}, 
\begin{eqnarray}
p_{\gb } (t) = \delta _{\gb , 0} +t_{\gb } +
\frac{1}{2!} \sum_{\hb ', \hb '' \in  {\bf Q} ;\hb '+\hb ''=\gb } t_{\hb 
'}
t_{\hb ''}
+ \frac{1}{3!} \sum_{\hb ', \hb '', \hb ''' \in {\bf Q} ;\hb '+\hb ''+\hb
'''=\gb
} 
t_{\hb '} t_{\hb ''} t_{\hb '''} +\cdots .\label{eqn;fortyseven}
\end{eqnarray}
\noindent
The set $\{ p_{\gb} \} _{ \gb \in {\bf Q}}$ 
defined above indeed consists of polynomials:
the sum in the r .h. s. of  (\ref{eqn;fortyseven})
is finite as there is a finite number of ways to present an element
of ${\bf Q}$ as a sum of elements of ${\bf Q}$ of positive degree.
It is also worth mentioning that one can grade the 
polynomial ring ${\bf C} [ t_{\hb} , \hb \in {\bf Q}]$
by setting $deg (t_{\hb }) = deg ( \hb )$. Under such assignment
$p _{\gb } (t) $ becomes a homogeneous 
polynomial of degree equal to $deg(\gb )$.

We will also need a generating function for the Schur polynomials. 
Let $z, \bar{z}$ be complex coordinates in the plane. 
We set $\zv ^{~ \gb } \equiv z^{a} \cdot \bar{z} ^{b}$, 
where $\gb =(a, b)$. Let also
\begin{eqnarray}
V(t,  ~ \zv ) = \sum_{\gb  \in  \Q} t_{\gb } \zv ^{~\gb
}.\label{eqn;fortysevena}
\end{eqnarray}
Then
\begin{eqnarray}
e^{V(t, ~ \zv) } =\sum_{\gb  \in \Q} p_{\gb } (t) \zv ^{~\gb
}.\label{eqn;fortyeight}
\end{eqnarray}
The generating function (\ref{eqn;fortysevena}) 
can be used to rewrite relations  
(\ref{eqn;fortyfive}) and (\ref{eqn;fortysix}) in a compact form. Let
\begin{eqnarray}
\f (\zv ) = \sum _{\gb \in {\bf G}} \f _{\gb } \zv ^{~\gb }
\label{freefield}
\end{eqnarray}
be a free fermion field operator.
Then a direct computation shows that
\begin{eqnarray}
e^{H(t)} \f (\zv ) e^{-H(t)} = e^{V (t, \zv )} \f (\zv ).
\label{evol}
\end{eqnarray}

To derive a free fermion representation
of the partition function of NMM we will need a pair
of operators called {\em fermionic projectors}:
\begin{eqnarray}
P _{+}= ~ : e^{\sum_{\gb  \in \Gm} \f_{\gb } \fb_{\gb }}:~,
\label{eqn;fortynine}\\
P _{-}= ~ : e^{-\sum_{\gb  \in \Gp} \f_{\gb } \fb_{\gb }}:~.
\label{eqn;fortyninea}
\end{eqnarray}
One can verify the following properties of fermionic projectors:
\begin{eqnarray}
P _{+} \fb _{\gb }=\f _{\gb } P _{+} =0, \mbox{ if } \gb  \in \Gm;
\label{eqn;fifty} \\
P _{-} \f _{\gb } = \fb _{\gb } P _{-} =0, \mbox{ if } \gb \in \Gp;
\label{eqn;fiftya} \\
P _{+} ^2 =P _{+} ~ ; ~ P _{-} ^2 = P _{-}.
\label{eqn;fiftyone}
\end{eqnarray}  
Eq.~(\ref{eqn;fiftyone}) is a direct consequence of (\ref{eqn;fifty}) 
and (\ref{eqn;fiftya}), which in turn
result
from the following equivalent representation
of fermionic projectors:
\begin{eqnarray*}
P _{+} = \prod_{\gb  \in \Gm} (1- \fb _{\gb } \f_{\gb }), \\
P _{-} =\prod_{\gb  \in \Gp}  (1- \f_{\gb } \fb _{\gb } ).
\end{eqnarray*}

Finally we are ready to introduce 
the main object of our interest.  Consider
the following fermionic correlator:
\begin{eqnarray}
\tau \left(U ,   A(\Phi ) | t\right)= \llu e^{H(t)} A(\Phi )   \ru,   
\label{eqn;fiftytwo}
\end{eqnarray}
where $U \subset \Q$ is a finite subset of $\Q \subset {\bf G}$,
\begin{eqnarray}
A(\Phi ) = ~ : e^{ \left\{\int_{{\bf C} ^2} dz d\bar{z} dw d \bar{w}~~
\Phi ( \zv ,   \wv ) \f _{+} ( \zv\,) \fb _{+} (\wv ^{-(1, 1)})
-\sum_{\gb  \in \Gp} \f_{\gb } \fb _{\gb } \right\}}: ~ ,   
\label{eqn;fiftythree}
\end{eqnarray}
 
\begin{eqnarray}
\f _{+} (\zv ) =\sum_{\gb  \in \Gp} 
 \f _{\gb } \zv ^{~\gb }~, ~~
\fb _{+} (\zv )=\sum_{\gb  \in \Gp } 
\fb _{\gb } \zv ^{~-\gb },
\label{eqn;fiftyfour}
\end{eqnarray}
and $\Phi$ is a real function (or a distribution) on ${\bf C} 
^2$. 
Our aim is to compute the correlation function (\ref{eqn;fiftytwo}). 
Consider the following matrix of an infinite size:
\begin{eqnarray}
A(\Phi )_{\gb, \hb} &=& 
\int_{{\bf C} ^2} dz d\bar{z} dw d \bar{w}~~
\Phi ( \zv ,   \wv ) \wv ^{~\gb} \zv ^{~\hb},
~~~~ \hb,~ \gb \in \Gp ,\nonumber \\
A(\Phi)_{\gb, \hb} &=& \delta_{\gb , \hb},~~~~\hb, ~\gb \in \Gm
\label{A}
\end{eqnarray}
and $A_{\gb, \hb} (\Phi) =0$ in all other cases.
One can verify the following set
of identities involving the matrix $A(\Phi )_{\gb, \hb}$:
\begin{eqnarray}
\sum_{\hb \in \G}  A(\Phi )_{\gb , \hb} \f _{\hb} A(\Phi ) 
= A(\Phi ) \f _{\gb} ,~ \gb \in \G .
\label{relation}
\end{eqnarray}
To prove (\ref{relation}) we consider separately cases
$\gb \in \Gm$ and $\gb \in \Gp$. In the former case
relation (\ref{relation}) reduces to  
$\f _{\gb} A (\Phi ) = A (\Phi ) \f_{\gb},~\gb \in \Gm$, which
is true, as the operator $A(\Phi )$ depends on fermionic
generators labeled by elements of $\Gp$ only. The latter
case follows in a straightforward way from the
representation
of the operator $A(\Phi )$ given below:
\begin{eqnarray}
A(\Phi ) = \sum _{m=0}^{\infty} \frac{1}{m!} 
\int dz_{1} d \bar{z}_{1} dw_{1} d
\bar{w}_{1}
\cdots  dz_{m} d \bar{z}_{m} dw_{m} d \bar{w}_{m} 
\Phi (\zv _{1} , \wv _{1} ) \cdots \Phi (\zv _{m} , \wv _{m} )  
\nonumber \\
\times \f _{+} (\zv _{1}) \cdots \f _{+} (\zv _{m} ) P_{-}
\fb _{+} (\wv _{m} ^{(-1 , -1)}) \cdots \fb _{+} ( \wv _{1} ^{(-1, -1)}).
\label{expansion}
\end{eqnarray}
In what follows we will always require the
non-degeneracy of 
the matrix $A_{\gb ,\hb} (\Phi)$ . We also
wish to remark that
operator
$A(\Phi )$ from (\ref{eqn;fiftythree}) is a counterpart of the operator used in
\cite{MMM} to
obtain a free fermion representation of 
conventional one-matrix models.

Using relations (\ref{relation}) and the fact that
$A(\Phi ) \rv =\rv$ and applying Wick's
theorem one can express 
$\tau ( U ,  A (\Phi)| t)$
in terms of two-point correlators: 
\begin{eqnarray}
\tau ( U ,   A (\Phi)| t)=det[ \lv \fb _{\gb } (-t) 
\sum_{\kb \in \Gp}  A(\Phi )_{\hb , \kb} \f _{\kb} 
 \rv]_{\gb , \hb \in U}, 
\label{eqn;fiftyfive} 
\end{eqnarray}
where
the determinant is taken with respect 
to some linear ordering of $U$.  Note however that
the change of order is a unitary 
transformation which does not change the determinant. 

It is not difficult to compute 
two-point correlation functions entering (\ref{eqn;fiftyfive}).
Substituting (\ref{A}) into (\ref{eqn;fiftyfive})
and using relation (\ref{evol})
we arrive at the following expression for $\tau (U, A (\Phi)| t)$:
\begin{eqnarray}
\tau (U ,  A (\Phi) | t) =det [Z _{\gb , \hb } (t)]_{\gb ,  \hb  \in U}, 
\label{eqn;fiftyseven} 
\end{eqnarray}
where
\begin{eqnarray}
Z_{\gb , \hb } (t) =\int_{{\bf C}^2 } dz d\bar{z}
dw d\bar{w}~~ \zv ^{~\gb } \wv ^{~\hb }
\Phi (\zv, \wv)e^{ V (t, \zv ) }
\label{eqn;fiftyeight}
\end{eqnarray} 
Suppose now that the set $U \subset {\bf G}$ and the function $\Phi$
which parametrize the 
$\tau$-function
are chosen to be the following:
\begin{eqnarray}
U_{N} = \biggl\{ \gb \in {\bf G} \bigg| \gb = (n, 0 ), n =0,1, \cdots N-1
\biggr\} ,
\label{eqn;sixty} \\
\Phi(\zv ,  \wv  )=\Phi _{NMM} (\zv ,  \wv  ) =
\delta (z - \bar{w} ) \cdot \delta
(\bar{z}
-w) \cdot e^{-| z | ^{2}}. 
\label{eqn;sixtya}
\end{eqnarray}
Substituting (\ref{eqn;sixty}) and (\ref{eqn;sixtya}) into
(\ref{eqn;fiftyseven}) and (\ref{eqn;fiftyeight}), we see that
\begin{eqnarray}
\tau ( U_{N} ,   A (\Phi _{NMM} ) | t ) =
det [ Z_{i, j} (t)]_{0 \leq i, j \leq N-1} ,\nonumber  \\
Z_{i, j} (t)=\int_{{\bf C}} dz d \bar{z}~ z^{i}
 \bar{z} ^{j} ~~e^{\sum_{m,   n \geq 0} (t_{(m, n)} - \delta _{m,1} 
\delta
_{n,1})  z^{m} \bar{z}
^n},\label{tau}   
\end{eqnarray} 
which coincides (up to nonessential factor) with the determinant formula
(\ref{eqn;five}) and   (\ref{eqn;six})
for the partition function of NMM with a  polynomial potential
equal to $- V(t, \zv) + | z | ^2$.  
Thus we proved that 
\begin{eqnarray}
Z_{N} = \frac{1}{N!} \tau \biggl( U_{N},   A(\Phi _{NMM}) \bigg|
t\biggr).\label{eqn;sixtyone}
\end{eqnarray}

To conclude the discussion of the present Section we would
like to mention that the partition functions of conventional
matrix models (such as Hermitian,  Unitary,  Complex one-matrix models
and Hermitian two-matrix models ) can be presented in the form
(\ref{eqn;fiftytwo})
under appropriate choices of $U$ and $\Phi$. 
To obtain the partition function of the 
hermitian matrix model, for example, set
\begin{eqnarray}
\Phi ( \zv ,  \wv ) =\Phi _{HMM} ( \zv ,  \wv )
 = \delta (z - \bar{w} ) \cdot
\delta ( \bar{z} - w ) \cdot 
\delta ( z -\bar{z} ) \cdot e^{- | z | ^{2} }
.\label{eqn;sixtythree}
\end{eqnarray}
Then
\begin{eqnarray}
\tau \biggl( U_{N} ,  A(\Phi _{HMM})  \bigg| t \biggr) 
= det [ Z_{i,j}]_{0 \leq i, j \leq N-1}
\label{eqn;sixtyfour}
\end{eqnarray}
with
\begin{eqnarray*}
Z_{i,j} = \int_{- \infty}^{\infty} dx ~ x^{i+j} 
e^{\Sigma_{m > 0} T_{m} x^{m}} \mbox{  and  }
T_{m}=\sum_{(k,l) \in  {\bf Q}}^{k+l=m}t_{(k,l)} - \delta _{m,1}. 
\end{eqnarray*}
Eq.~(\ref{eqn;sixtyfour}) is  
the determinant form of the partition function
of hermitian one-matrix model with the 
potential $- \sum_{p > 0} T_{p} x^{p}$. 

\section{ The Partition Function of NMM as a $\tau$-Function of 
the Extended-$KP(N)$ Hierarchy}

In this Section we are going to derive a system of differential
equations associated with correlation function (\ref{eqn;fiftytwo}).
In virtue of (\ref{eqn;sixtyone}) 
all results of the present Section apply as well to 
to the
partition function of NMM with an arbitrary polynomial potential.

The part of our construction dealing with free fermions
relies heavily on methods developed in \cite{JM1},
\cite{JM2} and \cite{JM3}, see \cite{JM} for a review. Our
consequent analysis of the emerging hierarchy of
differential equations shows that the original
approach of \cite{Sato} and \cite{Sato1} to the theory of 
KP equations can be extended to the multidimensional
case as well.

Let $U_{N}$ and $A$ be a subset of ${\bf Q}$ and an element
of the Clifford algebra given by (\ref{eqn;sixty}) and 
(\ref{eqn;fiftythree})
correspondingly (to simplify notations, from now on
we denote $A(\Phi )$ and $A(\Phi)_{\gb ,\hb }$
as $A$ and $A_{\gb ,\hb }$ respectively). 
 The following functions depending on $\zv$
and $t$'s are called  {\em  wave functions}:
\begin{eqnarray}
w_{\ppb } (\zv ,  t) =
\frac{\biggl\langle U_{N} \bigg| \fb _{N \cdot \ppb } e^{H(t)} \f (\zv ) A 
\bigg|U_N\biggr\rangle }
{\biggl\langle U_N\bigg| e^{H(t)} A \bigg|U_N\biggr\rangle} , 
\label{83}\\
w_{\qb } (\zv ,  t) = 
\frac{\biggl\langle U_{N} \bigg| \fb _{\qb } e^{H(t)} \f (\zv ) A 
\bigg|U_N\biggr\rangle }
{\biggl\langle U_N\bigg| e^{H(t)} A \bigg|U_N\biggr\rangle} ,
\label{84}
\end{eqnarray}
where $\f (\zv)  = \sum_{\gb \in {\bf G}} \f _{\gb} \zv ^{~\gb}$ is a free
field
operator; $\ppb = (1,0) \in {\bf G}, ~ \qb = (0,1) \in {\bf G}$. In our
notations for
the wave functions we
 suppress the dependence on $N$ and $A$ which are supposed
to be fixed. We assume that the common denominator in 
(\ref{83}) and (\ref{84}) is not equal to zero 
when all $t$'s are equal to $0$.
Then the wave functions make sense as formal power series
in $t$'s.
 
There is a linear relation imposed on the wave functions which follows
from the identity (\ref{relation}). Let us explore it.
Consider the Fourier decomposition
of wave functions  with respect to 
$\zv$:
\begin{eqnarray}
w_{\ppb } (\zv , t ) = \sum_{\gb  \in {\bf G}} w^{\gb }_{\ppb }
(t) \zv^{~\gb} ~,~ 
w_{\qb } (\zv , t ) = \sum_{\gb  \in {\bf G}} w^{\gb }_{\qb }
(t) \zv^{~\gb} 
\label{85}
\end{eqnarray}
It is easy to prove the following set of relations between
coefficients of such decomposition:
\begin{eqnarray}
\sum_{\hb  \in {\bf G}} A_{\gb , \hb } w^{\hb }_{\ppb } (t) =0 ~,~
\sum_{\hb  \in {\bf G}} A_{\gb , \hb } w^{\hb }_{\qb } (t) =0 
\mbox{ if } \gb  \in \Gm \bigcup U_{N} .
\label{86}
\end{eqnarray}
Here $A_{\gb , \hb }$ is a matrix defined by (\ref{A}).
To prove (\ref{86}) we observe that $w^{\hb} _{\ppb} (t)$
is proportional to
\begin{eqnarray*} 
\llun \fb _{N \cdot \ppb} e^{H(t)}  \f _{\hb} A \run .
\end{eqnarray*}
Multiplying it by $A_{\gb , \hb }$, summing
over $\hb $, and using (\ref{relation}) we get \hbox{$\llun \fb_{\ppb}
e^{H(t)} A \f
_{\gb } \run$,} 
which is zero if $\gb  \in \Gm \bigcup U_{N}$. The similar
arguments apply if we replace  $w^{\hb} _{\ppb} (t)$ with
 $w^{\hb} _{\qb} (t)$. 
Therefore,  (\ref{86}) is proven. 

Relations (\ref{86}) have their counterpart in the theory of KP
hierarchy (see e.g. \cite{JM} ) and are of prime importance
for our further considerations. Before one can make use
of them however, it is desirable  to rewrite equations (\ref{86})
in the form of  linear relations between the {\em finite} number
of unknowns. This is possible because of the following
representation of wave functions (\ref{83}) and (\ref{84}):
\begin{eqnarray}
w_{\ppb} (\zv, t) = \biggl(\zv ^{~N \cdot \ppb} + \sum_{\gb  \in U_{N}}
a_{\gb } (t) \zv^{~\gb } \biggr) e^{V(t, \zv \,)} ,
\label{87}\\
w_{\qb} (\zv, t) = \biggl(\zv ^{~ \qb} + \sum_{\gb  \in U_{N}}
b_{\gb } (t) \zv^{~\gb } \biggr) e^{V(t, \zv \,)},
\label{88}
\end{eqnarray}
where coefficients $\{ a_{\gb} (t), b_{\gb} (t) \} _{\gb \in U_{N}}$
depend on $t$'s and the choice of $A$. It is easy to express them
in terms of free fermion correlators, but we will not need
the explicit expressions. To verify (\ref{87}) one can perform the
following
computation: commute the field operator $\f (\zv )$ in the numerator
of (\ref{83}) with $e^{H(t)}$ using (\ref{evol}).
Then notice that 
\begin{eqnarray*}
\llun \fb _{N \cdot \ppb } \f _{\gb} =0,~ \gb \in \Gp \setminus 
\left( U_{N} \bigcup \{N \cdot \ppb \} \right)
\mbox{ and } \f _{\hb } e^{H(t)} A \run = 0 ,~ \hb \in \Gm,
\end{eqnarray*}
\noindent
in which the first equality
follows from the definition (\ref{eqn;forty})
of the generating vectors of $\lF$ while 
the other one is a consequence
of our choice of the operator 
$A$ defined in (\ref{eqn;fiftythree})
and the law (\ref{eqn;fortyfive}) 
of evolution of  fermionic generators:
\begin{eqnarray*}
\f _{\hb} e^{H(t)}  A \run =e^{H(t)} \sum _{\gb \in {\bf Q}} A 
\f_{\hb
- \gb}
p_{\gb} (-t) \run
=0 \mbox{, if } \hb \in \Gm .
\end{eqnarray*}
It only remains to show that the coefficient in front
of $\zv^{~N \cdot \ppb}$ in (\ref{87}) is indeed $1$.
But according to (\ref{83}) this coefficient is
equal to $\frac{\llun \fb _{N \cdot \ppb} \f _{N \cdot \ppb}
e^{H(t)} A \run}{\llun
e^{H(t)} A \run}$, which is $1$.  
Representation (\ref{87}) has been derived.
The derivation of (\ref{88}) can be performed
along the same lines.

Using the fact that $e^{V(t, \zv )} = \sum_{\gb \in {\bf G}} p_{\gb} (t) 
\zv ^{~\gb}$,
where we assumed that $p_{\gb} = 0$ if $\gb \in {\bf G} \setminus {\bf Q}$, 
and
comparing representations of wave functions
(\ref{85}) and ((\ref{87}), (\ref{88})) one can
rewrite the relations (\ref{86})
in terms of coefficients $\{ a_{\gb }, b_{\gb} \} _{\gb \in U_{N}}$:
\begin{eqnarray}
\sum_{\hb \in {\bf G}} A_{\gb, \hb} \left(p_{\hb - N \cdot \ppb}(t) +
\sum_{\kb \in U_{N}} a_{\kb} (t) p_{\hb - \kb} (t) \right) =0
\mbox{ if } \gb  \in \Gm \bigcup U_{N}, 
\label{89} \\
 \sum_{\hb \in {\bf G}} A_{\gb, \hb} \left(p_{\hb - \qb}(t) +
\sum_{\kb \in U_{N}} b_{\kb} (t) p_{\hb - \kb} (t) \right) =0 
\mbox{ if } \gb  \in \Gm \bigcup U_{N}. \label{eqn;seventynine}
\label{90}
\end{eqnarray}
Looking closer at relations (\ref{89}) and (\ref{90}) we see that the only
non-trivial ones
are those corresponding to $\gb \in U_{N}$. Indeed, if  $\gb \in \Gm$
then $A_{\gb , \hb }=0$ for $ \hb \in \Gp$ in accordance with
definition (\ref{A}). Thus  left hand
sides of relations (\ref{89}) and (\ref{90}) are just identical zeros if 
$\gb\in \Gm$. 

Let us consider now the following pair of differential operators:
\begin{eqnarray}
W_{\ppb} (t , \dv) &=& \dv ^{~N \cdot \ppb} +
 \sum_{\gb  \in U_{N}} a_{\gb } (t) \dv ^{~\gb }, 
\label{91}\\
W_{\qb} (t , \dv) &=& \dv ^{~ \qb} +
 \sum_{\gb  \in U_{N}} b_{\gb } (t) \dv ^{~\gb } ,
\label{92}
\end{eqnarray} 
which we call $wave$ $operators$.
Here $\dv ^{~\gb } \equiv \left( 
\frac{\partial}{\partial t_{\ppb}}\right)
^{g_{\ppb}}
\left( \frac{\partial}{\partial t_{\qb}}\right) ^{g_{\qb}}$
 and $\gb =g_{\ppb} \cdot \ppb + g_{\qb} \cdot \qb$ is
an arbitrary element of ${\bf G}$ decomposed in terms of
$\ppb$ and $\qb$. In what follows we will denote 
the complex variables
$t_{\ppb}$ and $t_{\qb}$
as $x$ and $y$, respectively. Thus $W_{\ppb}$ and  $W_{\qb}$
are polynomials in $\frac{\partial}{\partial x}$ and
$\frac{\partial}{\partial y}$ with coefficients depending
on ``times" $t$'s.
Evidently,
\begin{eqnarray}
w_{\ppb} (t, \zv )= W_{\ppb } (t, \dv ) e^{V(t, \zv)},~~~
w_{\qb } (t, \zv )= W_{\qb} (t, \dv ) e^{V(t, \zv)}.
\label{wW}
\end{eqnarray}
Noting also that $\dv ^{\gb} p_{\hb} (t) =
p_{\hb - \gb} (t)$ for $\gb \in {\bf Q}$, we can
rewrite relations (\ref{89}) and (\ref{90})
in terms of wave operators, which will
provide us with an 
alternative form
of equations (\ref{86}) suitable for our needs:
\begin{eqnarray}
W_{\ppb} h_{\gb} (t)=0=W_{\qb} h_{\gb} (t), ~ \gb \in U_{N},
\label{93}
\end{eqnarray}
where $h_{\gb} (t) = \sum_{\hb \in {\bf G}} A_{\gb, \hb} p_{\hb} (t)$.
Relations (\ref{93}) can be viewed as two $N \times N$ systems of
linear equations with respect to coefficients of wave operators.
Here is an explicit solution:
\begin{eqnarray}
a_{n \cdot \ppb} (t) =- \frac{det (\dv ^{0 \cdot \ppb} \vec{h} , \dv^{1 
\cdot
\ppb} \vec{h} ,
\cdots, \dv ^{(n-1) \cdot \ppb} \vec{h}, \dv ^{N \cdot \ppb} \vec{h}, \dv 
^{(n+1)
\cdot \ppb} \vec{h} ,
\cdots, \dv ^{(N-1) \cdot \ppb} \vec{h})}{det(\dv ^{0 \cdot \ppb} 
\vec{h},
\cdots , 
\dv ^{(N-1) \cdot \ppb} \vec{h})},  
\label{sola} \\
b_{n \cdot \ppb} (t) =- \frac{det (\dv ^{0 \cdot \ppb} \vec{h} , \dv^{1 
\cdot
\ppb} \vec{h} ,
\cdots, \dv ^{(n-1) \cdot \ppb} \vec{h} , \dv ^{\qb} \vec{h} , \dv ^{(n+1) 
\cdot
\ppb} \vec{h} ,
\cdots, \dv ^{(N-1) \cdot \ppb} \vec{h})}{det(\dv ^{0 \cdot \ppb} 
\vec{h},
\cdots , 
\dv ^{(N-1) \cdot \ppb} \vec{h})} ,  
\label{solb}
\end{eqnarray}
whre $n =0,1, \cdots N-1$ and $\vec{h}$ is a column
vector with elements  $ h_{\gb} , \gb \in U_{N} $.
It follows from the results of the previous Section that 
the common denominator of (\ref{sola}) and (\ref{solb})
is equal exactly to $\llun e^{H(t)} A \run$,
and is therefore an invertible power series in $t$'s.
Thus the wave operators are uniquely determined by conditions
(\ref{93}) and their coefficients are formal power series in ``times".

Let ${\bf B}$ be a ring of differential operators in
$\frac{\partial}{\partial x}$
and
$\frac{\partial}{\partial y}$ with coefficients being formal
series in variables $t$'s.
The statement bellow is a natural consequence of the formalism
developed:
\begin{lm}
Let $O \in {\bf B}$ be a differential operator such
that
\begin{eqnarray}  
O h_{\gb} =0,~ \gb \in U_{N} .
\end{eqnarray}
Then there are differential operators 
$b_{\ppb} \in {\bf B}$ and $b_{\qb} \in {\bf B}$ 
such that
\begin{eqnarray}
O = b_{\ppb} W_{\ppb} + b_{\qb} W_{\qb},
\end{eqnarray}
in other words $O =0 | _{mod ~ W_{\ppb} , W_{\qb}}$.
Moreover one can choose $b_{\ppb}$ to be of zeroth order in
$\frac{\partial}{\partial y}$.
\end{lm}

The proof of Lemma 1 is given in the Appendix.
The system of non-linear equations satisfied by coefficients of wave
operators is a direct consequence of the formulated statement. 
To see this we will differentiate each of the relations (\ref{93}) with
respect to
$t_{\hb}$ and use the fact that $\frac{\partial}{\partial t_{\hb}} 
h_{\gb}
(t)
= \dv ^{~\hb} h_{\gb} (t)$, a property of 
Schur polynomials. As a result we obtain the 
following set of identities:
\begin{eqnarray}
\left(\frac{\partial W_{\ppb ~ (\qb )} (t,  \dv )}{\partial t_{\hb}}
+ W_{\ppb ~ (\qb )} (t ,  \dv ) \dv ^{\hb} \right) 
h_{\gb}=0, \mbox{ if } \gb  \in U_{N} .
\label{96}
\end{eqnarray}
It also follows from (\ref{93}) that
\begin{eqnarray}
[ W_{\ppb} , W_{\qb} ] h_{\gb}=0 ,~ \gb \in U_{N}.
\label{97}
\end{eqnarray}
\noindent
Relations (\ref{96}) and (\ref{97}) state that operators
$\left(\frac{\partial W_{\ppb ~ (\qb )} (t,  \dv )}{\partial t_{\hb}}
+ W_{\ppb ~ (\qb )} (t ,  \dv ) \dv ^{~\hb} \right) \in {\bf B}$ and
$ [ W_{\ppb} , W_{\qb} ] \in {\bf B}$ annihilate
the set of functions $h_{\gb } ,~ \gb \in U_{N}$. Thus
it follows from Lemma 1 that 
\begin{eqnarray}
\frac{\partial W_{\ppb } (t,  \dv)}{\partial t_{\hb}}
+W_{\ppb } (t ,  \dv) \dv^{~\hb}  =0 
\bigg|_{mod ~ W_{\ppb} , W_{\qb}},~~ \hb \in {\bf Q}, 
\label{98} \\
\frac{\partial W_{\qb } (t,  \dv)}{\partial t_{\hb}}
+W_{\qb} (t ,  \dv) \dv^{~\hb}  =0 
\bigg|_{mod ~ W_{\ppb} , W_{\qb}} ,~~ \hb \in {\bf Q},
\label{99} 
\end{eqnarray}
\begin{eqnarray}
[ W_{\ppb} , W_{\qb} ] =Y \cdot W_{\ppb} = 0 \bigg| _{mod ~ W_{\ppb}} ,
\label{100}
\end{eqnarray}
where $Y \in \B$ is completely determined by $W_{\ppb}$ and $W_{\qb}$.
An explicit expression for the operator $Y$ in terms of wave
operators will be derived later and
the answer is given in (\ref{113a}).
Note that the r.h.s. of the (\ref{100}) is proportional to
$W_{\ppb}$ only,  which follows 
from the fact that  $[ W_{\ppb} , W _{\qb} ]$ is an operator in
$\frac{\partial}{\partial x}$ only.

For a fixed $N$, 
relations (\ref{98}) and (\ref{99})
 constitute the system of non-linear
differential equations for the unknown functions 
$\{ a_{n} (t) , b_{n} (t) \} _{n=0} ^{N-1}$  subject
to the constraint (\ref{100} ). More explicit form of this system will be 
presented later.

At this point we must comment on the correctness
of the definition (\ref{98}) and (\ref{99}) of equations of our hierarchy.
Depending on how we mode out the parts proportional to
$W_{\ppb}$ and $W_{\qb}$ we can obtain seemingly
different answers for the remainder. The reason for
such ambiguity is purely algebraic: 
the left $\B$-ideal ${\bf I} =
\biggl\{ O \in \B \bigg| O h_{\gb} =0 , \gb \in U_{N} \biggr\}$
consisting of all differential 
operators annihilating $h_{\gb}$ with $\gb \in
U_{N}$
is not freely generated by $W_{\ppb} , W_{\qb}$, there is the relation
(\ref{100}) between generators.
Moreover, this relation is the only one. In other words, we have
\begin{lm}
The left ideal $\I \subset \B$ can be described as follows in
terms of generators and relations:
\begin{eqnarray} 
{\bf I} = \biggl\langle  W_{\ppb} ,  W_{\qb} \Bigg| 
[ W_{\ppb} , W_{\qb} ] = 0 | _{mod ~ W_{\ppb}}  \biggr\rangle ,
\end{eqnarray}
which means that for any $O \in \I$ there are 
$b_{\ppb} \in \B$ and $b_{\qb} \in \B$
such that $O= b_{\ppb} W_{\ppb} + b_{\qb} W_{\qb}$. Moreover,
expressions $b_{\ppb} W_{\ppb} + b_{\qb} W_{\qb}$ and
$b'_{\ppb} W_{\ppb} + b'_{\qb} W_{\qb}$ are presentations
of the same element of $\I$ if and only if there is
an element $c \in \B$ such that
\begin{eqnarray*}
b_{\ppb} = b'_{\ppb} - c \cdot (W_{\qb} + Y ) \mbox{ and }
b_{\qb} = b'_{\qb} + c \cdot W_{\ppb} .
\end{eqnarray*}
Here $Y \in \B$ is an operator defined in (\ref{100}).  
{\em A proof of Lemma 2 is given in Appendix.}
\end{lm}
Therefore we conclude that all possible ways to write the
remainders in 
(\ref{98}) and (\ref{99}) lead to the same answer
if the relation (\ref{100}) is taken into account;
thus the the hierarchy
of equations we care about is completely determined by (\ref{98}),
(\ref{99}) and (\ref{100}).  

The practical way of deducing differential equations
from identities (\ref{98}), (\ref{99}) and 
(\ref{100}) can be extracted from the
proof of Lemma 1 in the Appendix. To illustrate
the result we present explicitly the simplest 
equations among (\ref{98}), (\ref{99}) and (\ref{100}) in the case when the 
set
$U_{N}$
consists of one point, i.e. $N=1$. The expressions
for the wave operators in this case are
\begin{eqnarray}
W_{\ppb  } =\frac{\partial}{\partial x}  + a (t),  
\label{101}\\
W_{\qb } =   \frac{\partial}{\partial y}  + b(t).  
\label{101a}
\end{eqnarray} 
The corresponding equations  for $\hb = (2,0) ,
(1,1)$
and  $(0,2)$ can be written as follows:
\begin{eqnarray}
\frac{\partial a}{\partial t_{(2,0)}} 
+ 2 (\partial _{x} a) a = \partial _{x} ^{2} a ,
\label{e1} \\
\frac{\partial b}{\partial t_{(2,0)}} 
+ 2 (\partial _{x} b) a = \partial _{x} ^{2} b; 
\label{e2} \\
\frac{\partial a}{\partial t_{(1,1)}} + (\partial _{x} a) b 
+ (\partial _{y} a) a = \partial _{x} \partial _{y} a ,
\label{e3} \\
\frac{\partial b}{\partial t_{(1,1)}} + (\partial _{x} b) b
+(\partial _{y} b) b =  \partial _{x} \partial _{y} b ;
\label{e4} \\
\frac{\partial a}{\partial t_{(0,2)}} 
+ 2 (\partial _{y} a) b = \partial _{y} ^{2} a ,
\label{e5} \\
\frac{\partial b}{\partial t_{(0,2)}} 
+ 2 (\partial _{y} b) b = \partial _{y} ^{2} b . 
\label{e6} 
\end{eqnarray}
The additional condition (\ref{100}) takes the form
\begin{eqnarray}
\partial _{y} a - \partial _{x} b =0 .
\label{e7}
\end{eqnarray}
Let us rewrite equations (\ref{e1}) - (\ref{e6}) 
and (\ref{e7})
 in terms of real variables and
real-valued unknown functions.
It follows from the reality condition 
imposed
on the potential $V(t, \zv )$ that $t _{(2,0)} = \overline{t_{(0,2)}}$,
$t _{(1,1)} = \overline{t_{(1,1)}}$  and 
$x= t _{(1,0)} = \overline{t_{(0,1)}} = \overline{y}$. Thus $a(t) =
\overline{b(t)}$
and we introduce the following new real variables:
\begin{eqnarray*}
v_{1}(t) \equiv a(t)+b(t)~,~ i v_{2} (t) \equiv b(t) -a(t), \\
r^{1} \equiv x+y~,~ i r^{2} \equiv x-y \\
\tau _{1}  \equiv t_{(1,1)}~,~ t_{(2,0)} \equiv 2(\tau _{2} + i \tau _{3}),   
t_{(2,0)} \equiv 2(\tau _{2} - i \tau _{3}).   
\end{eqnarray*}
Note that as far as transformation properties are concerned, $v_{i} (t)$  
with $i=1,2$,
is a $covector$ field (one-form). Relation
(\ref{e7}) states that this one-form is closed:
\begin{eqnarray}
d \left(v_{i} (t) d r^{i} \right) =0.
\label{102}
\end{eqnarray}
Equations (\ref{e1}) - (\ref{e6}) written in 
terms of $v_{i} (r, ~ \tau)$ acquire the form
\begin{eqnarray}
\frac{\partial \vec{v} _{\alpha} }{\partial 
t_{\alpha}} +(\vec{v}_{\alpha} \cdot \nabla ) 
 (\vec{v} _{\alpha} ) = \Delta _{g_{\alpha}} 
 \vec{v} _{\alpha},~ \alpha =1,2,3,
\label{103}
\end{eqnarray}
where  $\nabla = ( \frac{\partial }{\partial r^{1}},
\frac{\partial }{\partial
r^{2}} )$
is a gradient operator, $g_{\alpha}$'s are metric tensors,
$(g_{1}) _{ij} = \delta _{ij}, (g_{2}) _{ij} =\frac{1}{2} (\sigma _{3}) 
_{ij}, 
(g_{3}) _{ij} =(\sigma _{ 1} ) _{ij}$ and $\sigma _{3} = diag (1,-1)$,
$\sigma _{1} = antidiag (1,1)$. Matrices $\sigma$'s are the Pauli matrices.
An operator $\Delta _{g_{\alpha}} =g ^{ij} _{\alpha} 
\partial _{i} \partial _{j} $  is a Laplace
operator of two-dimensional space equipped 
with metric $g_{\alpha}$. Finally, 
$\vec{v} _{\alpha}(t)= \biggl( v^{1} _{\alpha}(t),
v^{2} _{\alpha}(t) \biggr)$ is a
vector field corresponding to a covector field 
$\biggl( v_{1} (t), v_{2} (t) \biggr)$ in the presence
of the metric $g_{\alpha}$, $v^{i} _{\alpha} = g^{ij} _{\alpha} v_{j}$. 
Note
that $g_{1}$ is a Euclidean metric,
while $g_{2}$ and $g_{3}$ are equivalent Minkowski metrics.

Relations (\ref{103}) for $\alpha =1$ (the Euclidean case) are called
two-dimensional
Burgers equations, \cite{Gurb1}.
As a result we see that two-dimensional Burgers equations are 
included
in an infinite hierarchy of non-linear differential equations, equations
(\ref{98})-(\ref{99}) and (\ref{100}) for
$N=1$. This hierarchy is completely
integrable in the sense 
that we can find all solutions in the class
of formal power series, see Theorem 1 below.\footnote{By the way, the
original (i.e. one-dimensional) Burgers 
equation \cite{Burgers} also can be included 
in the integrable hierarchy constituting a
certain reduction of KP hierarchy.
This statement can be easily extracted from the results of \cite{Ohta}. }

Burgers equations in one, two and
three spatial dimensions together with continuity equation and 
potentiality
condition (\ref{102})
can be used to model
potential turbulence. We refer the reader to 
\cite{Gurb} for a review and original
references. This book also describes an application of three-dimensional
analogue of equations (\ref{103}) to the study of large scale structure 
of the
Universe. The integrable structure we have discussed can prove useful
in the systematic analysis of the development of the turbulence in
the models based on Burgers equation: the integrable structure
of the Burgers hierarchy implies that the dynamics of the system
is constrained to the invariant subspaces of the phase
space (or `` state space" in the terminology of \cite{Landau}).
The addition of the small perturbation in the form of
the random force (see e.g. \cite{Sinai}) destroys integrability
and the system moves
towards chaos through the deterioration 
of  invariant subspaces in
accordance with Kolmogorov-Arnold-Moser theory.
The presented picture is very close to the existing
scenarios of the development of the turbulence
( Hopf-Landau scenario for example, see \cite{Landau} for
details) but
has a chance to admit a complete qualitative treatment.
Moreover, the integrability (in the sense of the presence
of integrals of motion) plays an important role in
the description of fully developed turbulence
of \cite{Polyakov} and \cite{Polyakov1}. For example,
Polyakov's anomaly introduced in \cite{Polyakov1}
is an anomaly of the conservation law. We hope to 
investigate the consequences of complete integrability
of Burgers hierarchy for the theory of Burgers
turbulence in the near future.

In the meantime let  us discuss the solutions to (\ref{103})  
subject to the condition (\ref{102}).
Solutions to (\ref{102}) which are defined at every 
point of the $(r^{1}, r^{2})$-plane  are of the following form:
\begin{eqnarray}
v _{i}  (t) = \partial _{j} \Psi (t),
\label{104}
\end{eqnarray}
where $\Psi (t)$ is an arbitrary function of $t$'s.
It is called a potential of the covector 
field $v _{i} (t)$. Comparing (\ref{104}) with (\ref{sola}) and (\ref{solb})
 we see that the whole class of solutions
to equations (\ref{103}) is given by
\begin{eqnarray} 
\Psi (t) =  ln \langle U_{1} | e^{H(t)}
A | U_{1} \rangle =ln \biggl\{ \int dz d \bar{z} [ \int dw d 
\bar{w} \Phi (\zv , \wv )] e^{V(t,\zv )} \biggr\}.
\end{eqnarray} 
We see that $\Psi (t)$ is a generalization of the Hopf-Cole 
(\cite{Hopf}, \cite{Cole})
solution to the Burgers equations (\ref{103}). 
The quantity inside the square brackets in the r.h.s.
of the above equation is determined by initial conditions.
The choice of $\Phi$ from (\ref{eqn;fiftythree}) in accordance with
(\ref{eqn;sixtya}) makes it
clear that the ($1 \times 1$) NMM solves the 
(2+1)-dimensional Burgers hierarchy.   
We also see that the
Hopf-Cole substitution 
\begin{eqnarray}
v_{i} (t) = \partial_{i} \Psi (t) = \partial _{i} ln (\tau (t)) 
\label{tf}
\end{eqnarray}
linearizes the (2+1)-d Burgers hierarchy,
 (\ref{98}) and (\ref{99}) at $N=1$, subject 
to the constraint (\ref{102}). Note that the function $\tau$ appearing
in (\ref{tf}) can be considered as a generating function for
the solutions of the (2+1)-d Burgers 
hierarchy and plays in this sense a role
of the so called $\tau$-function of this hierarchy, 
see \cite{SJM} where the notion of the 
$\tau$-function was introduced.      

Now we wish to analyze in some details the structure of
equations (\ref{98}), (\ref{99}) and (\ref{100}) for an arbitrary $N$. Their
equivalent
form is the following:
\begin{eqnarray}
\frac{\partial W_{\ppb } (t,  \dv)}{\partial t_{\hb}}
+W_{\ppb } (t ,  \dv) \dv^{\hb}  =O^{\hb}_{\ppb, \ppb} W_{\ppb}
 + O^{\hb}_{\ppb, \qb} W_{\qb},~  \hb \in {\bf Q}, 
\label{107}\\
\frac{\partial W_{\qb } (t,  \dv)}{\partial t_{\hb}}
+W_{\qb } (t ,  \dv) \dv^{\hb}  =O^{\hb}_{\qb, \ppb} W_{\ppb}
 + O^{\hb}_{\qb, \qb} W_{\qb},~  \hb \in {\bf Q}, 
\label{108}
\end{eqnarray}
\begin{eqnarray}
[W_{\ppb} , W_{\qb} ] = Y W_{\ppb} ,
\label{108a}
\end{eqnarray}
where $Y$ and $O _{ij}$ with $i,j = 
\ppb , \qb$ are elements of ${\bf B}$ and according to 
Lemma 1 we can choose these operators 
in such a way that $O _{i, \ppb}$ with 
$i=\ppb$ and $\qb$
are of zeroth order in $\frac{\partial}{\partial y}$. Such a choice permits
us to express the r.h.s. of (\ref{107})-( \ref{108a}) in terms of wave
operators alone.
To obtain such an expression we have to define first the right inverses
of the wave operators. A way to do this is the following. We set
\begin{eqnarray*}
W_{\ppb} ^{-1} = \partial _{x} ^{-N} \sum _{n \geq 0} 
d _{n}  \partial_{x} ^{-n}, \\
W_{\qb}^{-1} = \partial _{y} ^{-1} \sum_{n \geq 0} \tilde{e} _{n}
\partial_{y} ^{-n},
\end{eqnarray*}
where $\{ d _{n} \}_{n=0} ^{\infty} $
are formal power series in $t$'s, while   $\{ \tilde{e} _{n} 
\}_{n=0}
^{\infty} $ are differential operators in $\partial _{x}$ with
coefficients
being formal power series in $t$'s. Operators
$\{ \tilde{e} _{n} \}_{n=0}^{\infty} $
and series $\{ d _{n} \}_{n=0} ^{\infty} $
are uniquely determined by 
equations
\begin{eqnarray}
W_{\ppb} \cdot W_{\ppb} ^{-1} =1=W_{\qb} \cdot W_{\qb} ^{-1}.
\label{111}
\end{eqnarray}
The multiplication operation ``$\cdot$"
in the above equations is defined by the Leibnitz rule.
Relations (\ref{111}) considered as
equations with respect
to unknown quantities $\{ d _{n} \}_{n=0} ^{\infty} $
and $\{ \tilde{e} _{n} \}_{n=0}^{\infty} $ 
possess a unique solution.

Multiplying equations (\ref{107}),
and (\ref{108})
by $W_{\qb} ^{-1}$ from the right, extracting the differential parts 
and using the fact that 
the $O^{\hb}$'s are already differential operators
and the $O^{\hb}_{i, \ppb}$'s, $i = \ppb$ and $\qb $, are of 
zeroth order with respect to $\partial _{y} $ we find
\begin{eqnarray}
O^{\hb}_{i , \qb } = \biggl( W_{i} 
\dv ^{\hb} W_{\qb}^{-1} \biggr) _{+} ,~ i= \ppb \mbox{ and } \qb,
\label{112}
\end{eqnarray} 
where the subscript ``plus" denotes the operation 
of extracting the differential part of an operator, i.e.
if $O (t , \dv) =\sum_{\gb  \in {\bf G}}
c_{\gb } \dv ^{~\gb }$ then $\biggl( O (t,  \dv ) \biggr)_{+} =
\sum_{\gb  \in {\bf Q}} c_{\gb } \dv ^{~\gb }$. 

To compute operators $O^{\hb}_{\ppb, \ppb}$ and 
 $O^{\hb}_{\qb, \ppb}$ we substitute (\ref{112}) back into (\ref{107})
and (\ref{108}). Multiplying the resulting equations with $W_{\ppb}^{-1}$
and projecting  them onto differential 
operators in $\partial _{x}$ only we get:
\begin{eqnarray}
O^{\hb}_{i , \ppb} = \biggl( \bigl( W_{\ppb (\qb) } 
\dv ^{\hb} W_{\qb} ^{-1} \bigr) _{-} W_{\qb} W_{\ppb}^{-1} \biggr) _{(+,0)},~i=\ppb, \qb
\label{113}
\end{eqnarray}
where the subscripts denote the following operations:
for any pseudodifferential operator $O$, $O_{-} = O - O_{+}$
and $O_{(+,0)}$ is a projection of $O$ onto differential 
operators in $\partial _{x}$. 

It remains to compute the operator $Y$ entering the r.h.s. of 
(\ref{108a}).
This differential operator is of zeroth
order in $\partial _{y}$ and is
therefore equal to
\begin{eqnarray}
Y = \left([W_{\ppb} , W_{\qb}] W_{\ppb} ^{-1}\right)_{(+,0)}.
\label{113a}
\end{eqnarray} 
Substituting (\ref{112}) - (\ref{113a})
 into (\ref{107}) - (\ref{108a}) we obtain a system of 
equations  governing the evolution
of the wave operators:
\begin{eqnarray}
\frac{\partial W_{\ppb }}{\partial t_{\gb }} 
+ W_{\ppb }  \dv ^{\gb}=
 \left((W_{\ppb  } \dv ^{\gb} 
W_{\qb} ^{-1} )_{-} W_{\qb} W_{\ppb}^{-1} \right) _{(+,0)} W_{\ppb} 
+(W_{\ppb } \dv ^{\gb} W_{\qb}^{-1} ) _{+} W_{\qb} , ~\gb \in {\bf Q} 
\label{eightyfive} \\
\frac{\partial W_{\qb}}{\partial t_{\gb }} 
+ W_{\qb}  \dv ^{\gb}=
 \left((W_{\qb } \dv ^{\gb} 
W_{\qb} ^{-1} )_{-} W_{\qb} W_{\ppb}^{-1} \right) _{(+,0)} W_{\ppb} 
+(W_{\qb} \dv ^{\gb} W_{\qb}^{-1} ) _{+} W_{\qb} , ~\gb \in {\bf Q} 
\label{eightyfivea}
\end{eqnarray}
\begin{eqnarray}
[W_{\ppb} , W_{\qb} ] =\left([W_{\ppb} , W_{\qb}] W_{\ppb}
^{-1}\right)_{(+,0)} W_{\ppb} .
\label{114}
\end{eqnarray}
Equations (\ref{eightyfive}) and
(\ref{eightyfivea})
together with the constraint (\ref{114})
constitute a hierarchy of non-linear differential equations
which
can be viewed as a generalization of the
Sato equations
in the theory of KP equations (see \cite{Ohta} for review). Burgers 
equations
(\ref{103}) with the condition  (\ref{102}) give the simplest examples of
equations (\ref{eightyfive}) - (\ref{114})  for $N=1$.

For large enough $N$ equations (\ref{eightyfive})
 and
(\ref{eightyfivea})
contain first $n$ equations of KP hierarchy, $n << N$. To see
this, consider the equation (\ref{eightyfive}) for $W_{\ppb}$
when $\gb= n \cdot \ppb$ with $n > 1$. In order to present the answer
in the standard form we set $W_{\ppb} \equiv W$ and
$t _{n} \equiv t_{n \cdot \ppb}$ and obtain
\begin{eqnarray}
\frac{\partial W}{\partial t _{n}} + W \partial _{x} 
^{n}
= (W \partial _{x} ^{n} W ^{-1})_{+} W ,~ n > 1.
\label{115}
\end{eqnarray}
The set of equations (\ref{115}) constitutes
a certain reduction of KP hierarchy, 
which we describe as follows.
Take the solution to KP hierarchy (\cite{JM}),
\begin{eqnarray}
\tilde{W}=1 +w_{1} (t)\partial _{x} ^{-1} + w_{2} (t) \partial _{x} ^{-2} + 
\cdots ,
\label{116}
\end{eqnarray}
which satisfies an additional condition 
of $\tilde{W} \cdot \partial _{x} ^{N}$
being a differential operator. 
Then the operator $W=\tilde{W} 
\cdot \partial _{x}^{N}$ solves equations 
(\ref{115}). We 
call the hierarchy of equations (\ref{115}) the
$KP (N)$ hierarchy. This hierarchy has 
been described in details in \cite{Ohta}.

The KP hierarchy itself can be viewed
as a limit of $KP(N)$ as $N$ tends to infinity. Such a limit
makes sense due to the stabilization of equations in $KP(N)$ hierarchy:
its $n$-th equation is independent from $N$ if $n << N$.
However the hierarchy (\ref{115}) at finite $N$
can be of independent interest as well.
For instance, the exact version of the statement of the footnote 2
is that $KP(1)$
is an integrable system containing the (1+1)-d Burgers equation.
 
We can interpret equations (\ref{eightyfive}) and
(\ref{eightyfivea}) with (\ref{114})
as an
integrable extension of $KP (N)$ hierarchy 
(\ref{115}) to higher dimensions. Therefore,
we call such system the extended-$KP(N)$ hierarchy.
The $N=1$ example considered above supports such an interpretation:
the extended-$KP(1)$ hierarchy, or equivalently the (2+1)-d
Burgers hierarchy, is a  natural extension of the
$KP(1)$ hierarchy, or equivalently the
(1+1)-d Burgers
hierarchy.
  
Our main result about the extended-$KP(N)$ hierarchy is that it is a
$completely$ integrable extension of the $KP(N)$ hierarchy,
i.e. $all$ solutions to
the extended-$KP (N)$ hierarchy are of form
(\ref{sola}) and (\ref{solb}). To be
precise
we have the following theorem.
\begin{thm}
Let $\bf{C} [[t]]$ be a ring of formal power series in $t_{\gb}, \gb \in
{\bf Q}$, with
complex coefficients. Let $\bf{B}$ be a ring of differential operators in
$\partial _{x}$ and
$\partial _{y}$, with coefficients belonging to $\bf{C} [[t]]$. 
Operators $W_{\ppb}$ and $W_{\qb} \in \B$ of the form
\begin{eqnarray}
W_{\ppb} (t , \partial_{x} , \partial _{y} ) = \partial _{x} ^{N} +
 \sum_{n=1}^{N} a_{n} (t) \partial _{x} ^{n-1 }, 
\label{91z}\\
W_{\qb} (t ,\partial_{x} , \partial _{y}  ) = \partial _{y} +
 \sum_{n=1}^{N} b_{n } (t) \partial _{x} ^{n-1} ,
\label{92z}
\end{eqnarray} 
where $a_{n} (t)$ and $b_{n} (t) \in \bf{C} [[t]]$ for $n = 0, \cdots , N-1$, 
solve the system
(\ref{eightyfive}) - (\ref{114}) if and only if there 
exists a set $\{ h_{n} (t) \} _{n=1}^{N}$
of $N$ elements of $\bf{C} [[t]]$ such that 
\begin{eqnarray}
\mbox{The Wronskian } {\cal W}_{x} (h_{1} , \cdots , h_{N}) (t) 
\mbox{ of } h_{1}, \cdots, h_{N} \nonumber\\
\mbox{ with respect to the variable } x
\mbox{ is an invertible element of } \bf{C} [[t]], \label{133x}
\end{eqnarray}
\begin{eqnarray}
\vec{\partial} ^{\gb} h_{n} (t) = \frac{\partial}{\partial t_{\gb}} h_{n},~
n=1, \cdots, N;~ \gb \in \Q ,
\label{135x}
\end{eqnarray} 
and
\begin{eqnarray}
W_{\ppb} h_{n} (t)=0 = W_{\qb} h_{n} (t),~ n=1, \cdots , N,
\label{134x}
\end{eqnarray}

\end{thm}
  
Note that (i) operators $W_{\ppb}$ and $W_{\qb}$ defined in
(\ref{91z}) and (\ref{92z}) are the explicit form of
the wave operators defined in (\ref{91}) and (\ref{92}); 
(ii) the condition (\ref{133x}) implies in particular the 
linear independence of  $h_{1} (t) , 
\cdots , h_{N} (t) \in \bf{C} [[t]]$; 
(iii) relations (\ref{134x}) constitute a system
of linear algebraic equations for the coefficients $a_{1} (t),
\cdots , a_{N} (t)$ and $b_{1} (t), \cdots , b_{N} (t)$. This system
has a unique solution due to the linear independence of functions
$h_{1} (t) , \cdots , h_{N} (t)$.
Thus Theorem 1 indeed 
describes all solutions to (\ref{eightyfive}) - (\ref{114}) and
the explicit form of these solutions
is given by (\ref{sola}) and (\ref{solb}).

The proof of Theorem 1 is based on the following lemma. 
\begin{lm}
Let $W_{\ppb}$ and $W_{\qb}$ be any two elements
of $\bf{B}$ of the form (\ref{91z}) and
 (\ref{92z}) satisfying the relation (\ref{114}).
Then $dim(Ker W_{\ppb} \bigcap Ker W_{\qb})=N$
over the ring 
$\bf{C} [[\hat{t} _{\ppb} ,\hat{t} _{\qb}, t]]$ of formal
power series depending on $\{ t_{\gb} \mid
\gb \in \Q \setminus \{\ppb , \qb \} \}$.
Moreover the basis in $Ker W_{\ppb} \bigcap Ker W_{\qb}$ can be
chosen to satisfy condition (\ref{133x}).
\end{lm}
The proof of Lemma 3 is presented in Appendix.
However the proof of Theorem 1 based on 
Lemma 3 is so short and transparent
that we present it here. 
 
$\diamondsuit$ Suppose that operators 
$W_{\ppb}$ and $W_{\qb} \in \bf{B}$ solve the
hierarchy (\ref{eightyfive}) - (\ref{114}). Let us 
fix a basis $\tilde{h}_{1} (t), \cdots, \tilde{h}_{N} (t)$
of $KerW_{\ppb} \bigcap KerW_{\qb}$, which
satisfies condition (\ref{133x}). Lemma 3 states that
any element of $\bf{C} [[t]]$ annihilated by $W_{\ppb}$ and $W_{\qb}$
can be decomposed into linear combination of  $\tilde{h}_{1} (t), \cdots, 
\tilde{h}_{N} (t)$
with coefficients depending on all $t_{\gb}$'s, $\gb \in \Q \setminus 
\{ \ppb, \qb \}$. Let us apply operator 
equalities (\ref{eightyfive}) and (\ref{eightyfivea}) to
$\tilde{h}_{1} (t), \cdots, \tilde{h}_{N} (t)$. The r.h.s.'s of the 
results are identically $0$;
substracting $0 \equiv \frac{\partial (W_{\ppb (\qb )}\tilde{h}_{i} (t))}
{\partial t _{\gb}} $ from
the l.h.s.'s, we get:
\begin{eqnarray}
W_{\ppb  } D_{\gb} \tilde{h}_{i} (t) = 0=  
W_{\qb  } D_{\gb} \tilde{h}_{i} (t) ,~ \gb \in \Q , i= 1, \cdots N,
\end{eqnarray}
where $D_{\gb} \equiv \frac{\partial}{\partial t_{\gb}}- \dv ^{\gb}$.
Therefore $D_{\gb} \tilde{h}_{i} (t) \in KerW_{\ppb} \bigcap 
KerW_{\qb}$
for each $i$ and $\gb$. Then there exist $N \times N$ matrices
$A_{\gb}(t)$ with $\gb \in \Q$ independent  of $x$ and $y$ such that 
\begin{eqnarray}
D_{\gb} \tilde{h}_{i} (t) = \sum _{j} [A_{\gb}(t)]^{j}_{i} \tilde{h}_{j} 
(t).
\label{ker}
\end{eqnarray} 
These matrices are not unrelated. 
The fact that $[D_{\gb} , D_{\hb}]=0$ together
with the linear independence of
basic elements $\tilde{h}_{1} (t) , \cdots , \tilde{h}_{N} (t)$ 
yields
\begin{eqnarray}
\frac{\partial A_{\gb}}{\partial t_{\hb}} 
- \frac{\partial A_{\hb}}{\partial t_{\gb}}+
[A_{\gb}, A_{\hb}] =0,~ \gb ,~ \hb \in \Q 
\setminus \{ \ppb, \qb \}.
\label{zc}
\end{eqnarray}
These zero-curvature-like conditions 
imply that there is an $(x,y) $-independent
non-degenerate
matrix
$B (t)$, such that 
\begin{eqnarray}
A_{\gb} = \frac{\partial B}{\partial t_{\gb}}B^{-1} .
\label{flat}
\end{eqnarray}
Non-degeneracy of $B(t)$ means that $detB(t)$
is an invertible element of $\bf{C} [[t]]$.
Consider a new basis of  $KerW_{\ppb} \bigcap KerW_{\qb}$
defined by 
\begin{eqnarray}
h_{i} (t) = \sum_{j} B^{j}_{i} (t) \tilde{h}_{j} (t), ~ i= 1, \cdots, N.
\label{new}
\end{eqnarray}
Substituting (\ref{flat}) and (\ref{new}) into 
(\ref{ker}) we see that 
\begin{eqnarray}
D_{\gb} h_{i} (t) = 0, ~i= 1, \cdots, N.
\end{eqnarray}
Therefore the condition (\ref{135x}) of the theorem is satisfied
by elements $h_{1} (t) , \cdots, h_{1} (t)$.
The condition (\ref{133x}) is satisfied as well, since
${\cal W}_{x} (h_{1}, \cdots, h_{N}) (t) = detB(t) \cdot
{\cal W}_{x} (\tilde{h}_{1} , \cdots, \tilde{h}_{N})(t)$, and the r.h.s.
of this relation
is invertible in $\bf{C} [[t]]$.
Thus we have proven the ``if" part of the 
theorem (for each pair
$W_{\ppb}$ and $W_{\qb}$
solving (\ref{eightyfive}) - (\ref{114}), there 
exists a set of linearly independent
functions $h_{1} (t) , \cdots , h_{N} (t)$ satisfying
conditions (\ref{133x}) - (\ref{135x})). 

From our considerations which led to 
the hierarchy (\ref{eightyfive}) - (\ref{114})
we know that the inverse statement is also true:
any pair of operators $W_{\ppb}$ and $W_{\qb}$
annihilating $N$
functions that satisfy condition (\ref{133x}) and (\ref{135x}) 
solves the extended-$KP(N)$ hierarchy.
This concludes the proof of Theorem 1. $\diamondsuit$

Theorem 1 implies a geometric description of
the space of solutions to the 
extended-$KP(N)$ hierarchy.
Before we can give such a description
some additional notations are to be introduced.
Let ${\bf L} \subset {\bf C} [[t]]$
be a complex linear space consisting
of elements of ${\bf C} [[t]]$ which
are annihilated by operators 
$D_{\gb} \equiv \frac{\partial}
{\partial t_{\gb}}-\dv ^{~\gb},~ \gb \in \Q$. 
Consider a set of all $N$-dimensional linear
subspaces of ${\bf L}$ which
we identify with an infinite-dimensional
Grassmann manifold $Gr(\infty, N)$.
Let us remind that $Gr(\infty , N)$
is defined as a set of all $N$-dimensional
linear subspaces of ${\bf C} ^{\infty}$
(understood as a Tychonoff product),
see \cite{Fuchs} for details.
We equip $Gr(\infty , N)$ with the
structure of topological space
by declaring that the set ${\cal B}_{\gamma}$
consisting of $N$-dimensional linear
subspaces of $\bf{L}$ having non-degenerate
projections on the finite-dimensional linear
subspace $\gamma \subset {\bf C} [[t]]$ is open.
It follows from standard theorems of
analysis (see e. g. \cite{Kolm}) that 
the set ${\cal B} = 
\{ {\cal B}_{\gamma} \}_
{\gamma \subset {\bf C} [[t]]}$ 
together with an $\emptyset$
constitutes
a base for the topology on $Gr (\infty , N)$.  
Let $\biggl(Gr(\infty, N)\biggr) _{0}$ be an open subset
of $Gr(\infty, N)$ consisting of all
$N$-dimensional linear subspaces of
${\bf L}$ having a non-degenerate
projection on the subspace $\Pi$ of ${\bf C} [[t]]$
spanned by $\{ x^{n} \}_{n=0}^{N-1}$.
The space of solutions to the extended-$KP(N)$
hierarchy can be now described as follows.

\begin{cl} 
There is a one-to-one correspondence between
the set of solutions to 
(\ref{eightyfive}) - (\ref{114})
and points of $\biggl( Gr(\infty , N) \biggr) _{0}$.  
\end{cl}
$\diamondsuit$ First of all let
us prove that
two sets of functions
$h_{1} (t) , \cdots , h_{N} (t)$ and 
$\tilde{h}_{1} (t) , \cdots , \tilde{h}_{N} (t)$
both satisfying (\ref{133x}) and (\ref{135x})
determine the same pair of operators $W_{\ppb}$
and $W_{\qb}$ from (\ref{91z}) and (\ref{92z})
which annihilate them if and only if
these two  sets of functions are related by a constant non-degenerate
linear transformation.

The ``if" statement is a direct consequence  of the Kramer's formula (see
(\ref{sola}) and (\ref{solb})).
Conversely, suppose that  $W_{\ppb}$ and $W_{\qb}$
annihilate two sets of functions
$h_{1} (t) , \cdots , h_{N} (t)$ 
and $\tilde{h}_{1} (t) , \cdots , \tilde{h}_{N} (t)$
satisfying (\ref{133x}) and (\ref{135x}).
Each of these sets span $KerW_{\ppb} \bigcap KerW_{\qb}$.
Therefore, there is a non-degenerate $N \times N$ matrix $M(t)$
independent of $x,y$ such that 
$h_{i} (t) = \sum_{j} M^{j}_{i} (t) \tilde{h}_{j} (t)$. 
Applying the operator
$D_{\gb}$ to both sides of the last equality and using the fact that
$D_{\gb} h_{i} (t) = D_{\gb} \tilde{h}_{i} (t) =0, i=1, \cdots , N$, and 
the linear
independence of elements 
$\tilde{h}_{1},\cdots, \tilde{h}_{N}$, we get 
$\frac{\partial M(t)}{\partial t_{\gb}}=0, \gb \in \Q$. Therefore,
$M$ is a constant non-degenerate matrix.
Thus any two sets
of $N$ elements of $\bf{C} [[t]]$ obeying (\ref{133x}) and
(\ref{135x}) satisfy (\ref{134x}) 
iff they are related by a constant non-degenerate linear
transformation.

Theorem 1 implies that we have just
constructed a
one-to-one map from the space of 
solutions 
to the extended-$KP(N)$ hierarchy
to the set of $N$-dimensional
linear subspaces of ${\bf L}$ or,
equivalently, $Gr(\infty, N)$.
This map is not onto: a linear
subspace of ${\bf L}$ belongs
to the image iff one can find
a basis $\{ h_{n} \}_{n=1}^{N}$
of this subspace such that
(\ref{133x}) is satisfied. But
this condition is equivalent
to the non-degeneracy of the projection
of our subspace onto the subspace
$\Pi$ described above. Thus the image
of the map in question is
exactly $(Gr(\infty, N))_{0}$ and
the Corollary is proved. $\diamondsuit$

Finally, let us discuss a relation between the solutions (\ref{sola})
and (\ref{solb}) to
the extended-$KP(N)$
hierarchy and the partition function
of NMM or, more generally, the fermionic correlators
(\ref{eqn;fiftytwo})
with $U= U_{N}$.
Consider the so called $vertex ~ operator$:
\begin{eqnarray} 
\X =e^{V (t ,  \zv )} e^{-V(\frac{1}{n} \dq,~ \zv ^{~-\ppb })}, 
\label{vertex}
\end{eqnarray}
where $V (\frac{1}{n} \cdot \dq , \zv ^{~-\ppb} ) =\sum _{n > 0} \frac{1}{n}
\frac{\partial}{\partial t_{n \cdot \ppb}} \zv ^{~- n \cdot \ppb}$. 
It is proved in the Appendix that the wave-function $w_{\ppb}$ which encodes
half of the solution to the extended-$KP(N)$ hierarchy can be expressed
through the fermionic correlator:
\begin{eqnarray}
w_{\ppb} (t, \zv ) = \frac{ X(t,  \zv \,) \tau (U_{N}, A, t)}
{\tau (U_{N} , A,  t)} \label{eqn;ninetynine}.
\end{eqnarray} 
This equation can be viewed as a generalization
of previously  known one-dimensional bosoni\-zation formulae  
\cite{JM} .
Knowing the fermionic correlator (\ref{eqn;fiftytwo})
we can compute $w_{\ppb} (\zv ,t)$ or, equivalently
the set of functions $\{ a_{n} (t) \} _{n=0}^{N-1}$.
The other half of the solution is given by
the set $\{ b_{n} (t) \} _{n=0}^{N-1} $. 

Even though we do not have
at the moment explicit formulae for the functions $b(t)$'s in terms
of the fermionic correlators (\ref{eqn;fiftytwo})
we believe that functions $b(t)$'s are also
determined by $\tau (U_{N}, A , t)$. The reason for such a 
conjecture
is very simple: all solutions at hand are determined completely by
the matrix $A_{\gb , \hb}$ with $\gb \in U_{N}$ 
and  $\hb \in {\bf Q}$ which in turn
can be restored from $\tau (U_{N} , A , t)$. This conjecture
has been verified in
case $N=1$, see (\ref{104}).
So we conclude that fermionic correlators $\tau (U_{N}, A , t)$
with operators $A$ from (\ref{eqn;fiftythree})
play a role of $\tau$-functions
for the solutions (\ref{sola}) 
to the extended-$KP(N)$ hierarchy.

We hope to continue the investigation of the structure of 
hierarchies (\ref{eightyfive}) - (\ref{114}). One of the most important
questions to be answered here
is following: is there a universal integrable system from
which one can obtain all 
hierarchies (\ref{eightyfive}) and
(\ref{eightyfivea}) for $N=1 , 2 , \cdots$ by means of
reductions? The answer to
this question
is not clear. The reason is that equations  (\ref{eightyfive}) and
(\ref{eightyfivea}) of 
the extended-$KP(N)$ hierarchy 
{\em do not} stabilize as $N$ becomes larger. This
is readily seen from the fact that the number of
elements of $\Q$ of a given degree $d$ grows
with $d$.
Thus the relation between extended-$KP(N)$ hierarchies
and
this hypothetical universal structure should be different
from the known relation between $KP(N)$ hierarchies and $KP$ hierarchy.

\section{ Ward identities in the NMM }

It is well-known (see \cite{Mironov}, 
\cite{Ambjorn}, \cite{Itoyama}, see \cite{Morosov}
for review) that
partition functions of hermitian, unitary, complex, etc.,
matrix models exhibit invariance with respect to
a subalgebra of an algebra of holomorphic diffeomorphisms
of a complex plane,  or Virasoro algebra with zero
central charge. This invariance can be presented
in the form of
the so called Virasoro constraints imposed on
the partition function of a matrix model. 
It is also known that Virasoro constraints can be rewritten
in the form of
loop equations, \cite{Paffuti}, 
\cite{Kazakov}, see \cite{Mak2}
for a review.
These equations happen to be exactly solvable
in certain scaling limits thus providing a powerful tool
for a study of matrix
models.

The aim of the present Section is to demonstrate that 
the partition function of NMM is also subject
to an infinite set of constraints.  These constraints 
generate
a  subalgebra of  $\w$ algebra of all infinitesimal diffeomorphisms
of the complex plane.   

We start with some heuristic considerations intended to unveil 
the reasons for the appearance of a subalgebra of the $\w$ algebra
in the normal matrix model. 

Consider a family of $0$-dimensional field theories
with action $V$ parametrized by the set
of coupling constants $\{ t _{kl} \} _{k ,l \geq 0}$, where
$t_{k,l}= \overline{t _{l,k}}$:
\begin{eqnarray}
V= \sum_{i=1}^{N} \sum_{k, l \geq 0} t_{kl} z^{k} _{i}    
\bar{z} ^{l} _{i}.\label{eqn;onehundred} 
\end{eqnarray}
This family is equivalent in the field-theoretical framework to the
NMM itself as for each fixed set of coupling constants
(\ref{eqn;onehundred}) gives a NMM potential entering (\ref{eqn;four}). 
The algebra of the following reparametrizations acts in the space
of field theories (\ref{eqn;onehundred}):
\begin{eqnarray}
z_{i} \rightarrow z_{i} + \epsilon ~ z_{i} ^{m+1} \bar{z} _{i} ^{n}
\label{eqn;onehundredone}\\
\bar{z} _{i} \rightarrow \bar{z} _{i} +\bar{\epsilon} ~ \bar{z} _{i} ^{m+1}
z_{i} ^{n}, \label{eqn;onehundredtwo}
\end{eqnarray}
where $m,n \geq 0$. 
These reparametrizations can be presented as vector fields 
on the parameter space of NMM with the potential (\ref{eqn;onehundred}):
\begin{eqnarray}
\delta _{m,n} V = \epsilon ~ w_{m,n} V 
+ \bar{\epsilon} ~ \overline{w}_{m,n} V,
\label{eqn;onehundredthree}
\end{eqnarray} 
where
\begin{eqnarray}
w_{m,n} = \sum _{k, l \geq 0} k ~ t_{kl} 
\frac{\partial}{\partial t_{k+m,  l+n}}, ~~
\overline{w} _{m,n } = \sum _{k, l \geq 0} l ~ 
t_{kl} \frac{\partial}{\partial t_{k+n, l
+m}}.\label{eqn;onehundredfour}
\end{eqnarray}
These vector fields obey the following commutation relations:
\begin{eqnarray}
[w_{m,n},w_{p,q} ] &=& (p-m) w_{m+p,n+q}, \nonumber \\
\mbox{[} \overline{w} _{m,n},\overline{w}_{p,q} ] 
&=& (p-m)\overline{w}_{m+p,n+q}, 
\label{eqn;onehundredfive} \\
\mbox{[} w_{m,n},\overline{w}_{p,q} ] &=&
 q \cdot\overline{w}_{n+p,m+q} -
n \cdot w_{m+q, n+p}, \nonumber 
\end{eqnarray}
in which we recognize the relations for the $\w  $ algebra generated
by operators $z^{n+1} \bar{z} ^{\,m} \frac{\partial}{\partial z}$ 
and $\bar{z} ^{\,n+1} z^{n} \frac{\partial}{\partial \bar{z}}$. 
Operators (\ref{eqn;onehundredfour}) span
a subalgebra of $\w $ consisting of  all
polynomial reparametrization of the plane. 
So we conclude that the subalgebra of 
$\w $ describes classical
symmetries of NMM.\footnote{The reasonings above
are inspired by \cite{Mironov}. }

Having discovered the 
 $\w$ symmetry on the ``classical" level we gain the hope
that it translates to the
``quantum" level in the form of corresponding constraints
(Ward identities) on the partition function (\ref{eqn;one}). 
To see this directly let
us perform the change of variables given by
(\ref{eqn;onehundredone}),
(\ref{eqn;onehundredtwo}) in the integral (\ref{eqn;four}). 
As a consequence
of the fact that the integral doesn't depend on the choice
of the integration variables we obtain the following
set of identities:
\begin{eqnarray}
 \int \prod_{i=1}^{N} dz_{i} d \bar{z} _{i}
| \Delta (z) | ^2 e^{-tr \sum_{k, l \geq 0} t_{kl} M^{k} \overline{M} ^{l}}
\left(- \sum_{p,q \geq 0} p t_{pq} tr (M^{m+p} \overline{M}) ^{n +q}\right. \nonumber\\
+\frac{1}{2} (m+1) tr(M ^{m} \overline{M} ^{n}) +
\frac{1}{2} \sum_{p=0}^{m} tr (M^{m-p} \overline{M} ^{n}) trM^{p}  \nonumber \\
+\left.\frac{1}{2} \sum_{p=0}^{n-1} \sum_{k \neq l}^{N} \bar{z} _{k} ^{n-1-p}
\bar{z} _{l} ^{p} z_{l} ^{m+1} 
\frac{\bar{z} _{k} - \bar{z} _{l}}{z_{k} -z_{l} } \right)    =0,
\label{eqn;onehundredsix}
\end{eqnarray}
and its complex conjugate.
Here $M$ and $\overline{M}$ are diagonal matrices with 
$M_{ii} =z_{i},  \overline{M} _{ii} =\bar{z} _{i}$. 
Our goal is to rewrite (\ref{eqn;onehundredsix}) in the form of 
differential
constraints  applied
to the partition function itself.  The obvious obstacle for doing so is
the term with double summation 
in the left hand side of (\ref{eqn;onehundredsix}).  To
overcome
it we present a fraction 
$\frac{\bar{z} _{k} -\bar{z} _{l}}{z_{k} -z_{l}}$ 
in the following form:
\begin{eqnarray}
\frac{\bar{z} _{k} -\bar{z} _{l}}{z_{k} -z_{l}}
=(\bar{z} _{k} - \bar{z} _{l} )^{2} \int_{0}^{\infty}
d \omega e^{- \omega (| z_{k} -z_{l} | ^2 +\eta)},
\label{eqn;onehundredseven}
\end{eqnarray}
where $\eta > 0$ is an infinitesimal parameter. 
Note that if $z_{k} =z_{l}$ the right hand side of
(\ref{eqn;onehundredseven}) is
equal to $\frac{0}{\eta}=0$,  therefore (\ref{eqn;onehundredseven})
can be considered as a continuation of the fraction at hands
to all values of $z_{k}, z_{l}$. 
The replacement of $\frac{\bar{z}_{k} -\bar{z} _{k} }{z_{k} -z_{l}}$
with the integral (\ref{eqn;onehundredseven}) in 
(\ref{eqn;onehundredsix})
alters the integrand on the set of measure $0$
and doesn't change the integral itself. Using this remark we can
substitute (\ref{eqn;onehundredseven}) into (\ref{eqn;onehundredsix}) to
arrive at the
desired result:
\begin{eqnarray}
W_{mn} Z_{N} =0 ~;~\overline{W}_{mn} Z_{N} =0 ,
\label{eqn;onehundredeight}
\end{eqnarray}
where $m,n \geq 0$ and
\begin{eqnarray}
W_{m,n} &=& \sum_{k, l \geq 0} k ~ t_{kl} 
\frac{\partial}{\partial t_{m+k, n+l}}+ \biggl\{ -
\frac{1}{2} (m+1) \frac{\partial}{\partial t_{m, n}} \nonumber \\
&+& \frac{1}{2} \sum_{p=0}^{m} \frac{\partial}{\partial t_{m-p, n}}
\frac{\partial}{\partial t_{p, 0}}   
+\frac{1}{2} \sum_{p=0}^{n-1} 
\int_{0}^{\infty} d \omega
e^{-\epsilon \cdot \omega}   
\biggl( \frac{\partial}{\partial t_{0, n+1-p}} e^{-\omega \hat{D} }
\frac{\partial}{\partial t_{m+1, p}}   \nonumber \\
&+& \frac{\partial}{\partial t_{0, n-1-p}}
e^{-\omega \hat{D} }
\frac{\partial}{\partial t_{m+1, p+2}}
-2  \frac{\partial}{\partial t_{0, n-p}} e^{-\omega \hat{D} }
\frac{\partial}{\partial t_{m+1, p+1}} \biggr) \biggr\} ,   
\label{eqn;onehundrednine}
\end{eqnarray}
\begin{eqnarray}
\overline{W}_{m,n} &=& \sum_{k, l \geq 0} l~ t_{kl} 
\frac{\partial}{\partial t_{n+k, 
m+l}}+ \biggl\{
- \frac{1}{2} (m+1) \frac{\partial}{\partial t_{n, m}} \nonumber \\
&+& \frac{1}{2} \sum_{p=0}^{m} \frac{\partial}{\partial t_{n, m-p}}
\frac{\partial}{\partial t_{0, p}}  
+\frac{1}{2} \sum_{p=0}^{n-1} 
\int_{0}^{\infty} d \omega
e^{-\epsilon \cdot \omega} 
 \times \biggl( \frac{\partial}{\partial t_{n+1-p, 0}} e^{-\omega \hat{D} }
\frac{\partial}{\partial t_{p, m+1}} \nonumber \\
&+& \frac{\partial}{\partial t_{n-1-p, 0}}
e^{-\omega \hat{D} }\frac{\partial}{\partial t_{p+2, m+1}}
-2  \frac{\partial}{\partial t_{n-p, 0}} e^{-\omega \hat{D} }
\frac{\partial}{\partial t_{p+1, m+1}} \biggr) \biggr\}. 
\label{eqn;onehundredninea}
\end{eqnarray}
Here 
\begin{eqnarray}
\hat{D} \equiv \stackrel{\leftarrow}{\delta} 
\stackrel{\leftarrow}{\bar{\delta}} -
\stackrel{\rightarrow}{\delta} \stackrel{\leftarrow}{\bar{\delta}} -
\stackrel{\leftarrow}{\delta} \stackrel{\rightarrow}{\bar{\delta}} +
\stackrel{\rightarrow}{\delta}
\stackrel{\rightarrow}{\bar{\delta}}, \label{DS}
\end{eqnarray} 
and $\delta$ and $ \bar{\delta}$ are the shift operators defined as follows:
\begin{eqnarray}
\delta (\frac{\partial}{\partial t_{m, n}}) 
\equiv \frac{\partial}{\partial t_{m+1, n}}, \label{eqn;onehundredtena}\\
\bar{\delta} (\frac{\partial}{\partial t_{m, n}}) 
\equiv \frac{\partial}{\partial t_{m, n+1}}.\label{eqn;onehundredtenb} 
\end{eqnarray}
The arrows above operators $\delta,  \bar{\delta}$ 
in  (\ref{DS}) indicate the direction
in which these operators apply. 
The operators $e^{-\omega \hat{D}}$ in (\ref{eqn;onehundrednine})
and (\ref{eqn;onehundredninea})
act within 
the brackets encompassing them.  Note that the first 
terms in the expressions (\ref{eqn;onehundrednine}) and
(\ref{eqn;onehundredninea}) for the operators
$W_{mn}$ and $\overline{W} _{mn}$ coincide with the ``classical"
$\w $ generators (\ref{eqn;onehundredfour}). The additional ``anomalous" 
terms in the curly brackets of expressions for 
$W_{mn}$ and $\overline{W} _{mn}$
appear from the variation of the measure in the 
integral (\ref{eqn;four}). 

One can check directly that operators 
(\ref{eqn;onehundrednine}) 
and (\ref{eqn;onehundredninea}) 
satisfy the commutation relations
(\ref{eqn;onehundredfive}), thus providing us
with the $\w $ constraints for the NMM. 
It follows from (\ref{eqn;onehundredfive}) 
that
operators $W_{m, 0}$ and $\overline{W} _{n, 0}$
generate two commuting copies of the 
subalgebra of Virasoro algebra. Thus the partition
function of NMM model
obeys Virasoro constraints as well. 
There is a nice feature of Virasoro
generators $W_{m, 0}$ and $\overline{W }_{n, 0}$
following from their representation
(\ref{eqn;onehundrednine}) and
(\ref{eqn;onehundredninea})
: the tedious terms involving integrals over
$\omega$
are absent in this case,  so the expressions 
for Virasoro generators $W_{m, 0}$
and $\overline{W }_{n, 0}$ are similar to the ones
appearing in the conventional matrix models (cf.  \cite{Morosov}):
\begin{eqnarray}
W_{m,0} = \sum_{k, l \geq 0} k ~t_{kl} \frac{\partial}{\partial t_{m+k, l}} -
\frac{1}{2} (m+1) \frac{\partial}{\partial t_{m, 0}}
+ \frac{1}{2} \sum_{p=0}^{m} \frac{\partial}{\partial t_{m-p, 0}}
\frac{\partial}{\partial t_{p, 0}}, 
\label{vir} \\
\overline{W}_{n,0} = \sum_{k, l \geq 0} l~t_{kl} \frac{\partial}{\partial t_{k, 
n+l}}
- \frac{1}{2} (n+1) \frac{\partial}{\partial t_{0, m}}
+ \frac{1}{2} \sum_{p=0}^{n} \frac{\partial}{\partial t_{0, n-p}}
\frac{\partial}{\partial t_{0, p}}.   
\label{vira}
\end{eqnarray} 
 
The generators of $\w $ given by (\ref{eqn;onehundredone})
and  (\ref{eqn;onehundredtwo}) can
be presented in the form
$ \bar{z}^{n} l _{m}$ and 
$\overline{w }_{m,n}= z^{n} \overline{l} _{m}$,
where $l_{n},  \overline{l }_{m}$ generate
two commuting copies of Virasoro algebra. 
In this representation
the invariance with respect 
to both Virasoro algebras implies $\w $-invariance. 
One might hope that, correspondingly, the invariance
of the partition function with respect to $W_{m,0}$
and $\overline{W}_{n,0}$ of equations (\ref{vir}) and (\ref{vira})
implies the invariance with respect to $\w$ generated
by (\ref{eqn;onehundrednine}) and
(\ref{eqn;onehundredninea}). 
Presently we do not have any results which
could justify such a hope.

In conclusion of the present Section we would like to 
make the following comment. 
Usually the derivation of Ward identities in the conventional
matrix models deals with the partition function written as an 
integral over the subset of the set of all matrices. 
Unfortunately, such a
representation of the partition function of the NMM
is unknown due to the non-linearity of the space
of normal matrices.  
However, we have managed to derive $\w$-constraints
for NMM starting from
its the eigenvalue form (\ref{eqn;four}).  
Actually, this approach has its own advantages.
For example we are now able
to apply similar considerations to
analyze the structure of Ward
identities in the models which are genuinely non-matrix.   

For example, consider a
one-dimensional $\beta$-{\em model}, which
was originally introduced in \cite{Dyson}, 
see \cite{Mehta} for
a review. This and similar models  reappeared in recent literature
in quite different contexts such as 
study of chaotic systems \cite{Altshuler} and fractional statistics
\cite{Semenoff}, \cite{Semenoff1}. The partition function of the model is
\begin{eqnarray}
Z_{N} (\beta ,  t) = \int \prod _{i=1}^{N} dx_{i} | \Delta (x) | ^{\beta} 
e^{-\sum_{j=1}^{N} \sum_{k=1}^{\infty} t_{k} x_{j}
^{k}},\label{eqn;onehundredeleven} 
\end{eqnarray}
where $\beta$ is any non-negative number and the integration
goes over ${\bf R} ^N$. 
The orthogonal,  hermitian and quaternionic matrix models
are the particular cases of the $\beta$-model
for the values of $\beta$ equal to $1, 2$ and $4$ correspondingly. 
In general (\ref{eqn;onehundredeleven}) can be thought of as a classical
partition
of the two-dimensional Coulomb beads threaded on a string;
$\beta$ is identified in this case with the charge of a bead.  The
expression in the exponent has a meaning of the external
potential. 
Exploiting the invariance of the integral (\ref{eqn;onehundredeleven})  
under
the change
of variables $x_{i} \rightarrow x_{i} + \epsilon x_{i} ^{n+1}, 
n \geq -1, $
we obtain the following set of Virasoro constraints
satisfied by the model:
\begin{eqnarray}
L^{\beta} _{n} Z_{N} (\beta , t) =0,  n\geq -1, 
\label{eqn;onehundredtwelve} 
\end{eqnarray} 
\begin{eqnarray}
L^{\beta} _{n}  \equiv \frac{\beta}{2} 
\sum_{k=0}^{n} \frac{\partial}{\partial t_{k}}
\frac{\partial}{\partial t_{n-k}} - 
(1-\frac{\beta}{2}) (n+1) \frac{\partial}{\partial t_{n}}+
\sum_{k=1}^{\infty} k t_{k} \frac{\partial}{\partial
t_{k+n}}.\label{eqn;onehundredthirteen}
\end{eqnarray}
 
To our best knowledge this simple observation was not done before. 

\section{Conclusion}

We see that NMM is a highly tractable yet nontrivial model  possessing 
rich structures. 
The correlation functions of NMM  with an axially symmetric 
potential can be expressed in terms of a holomorphic  function of one
variable. This holomorphic property
leads
to the universality of the correlation functions in the scaling limit.
In the particular
case of monomial potentials this holomorphic function is expressible
in terms of degenerate hypergeometric functions. The
NMM admits
a free fermion representation. Using it we find that 
the behavior of the partition function of the NMM  with respect to an
arbitrary variation
of the potential is governed by a completely
 integrable system of non-linear
differential equations.
This integrable system constitutes a 
multidimensional extension of 
the $KP(N)$ 
hierarchy. In the simplest case when $N=1$
it contains (2+1)-d Burgers
equations. The partition function
of the NMM is subject to $\w$-constraints
which reflect the symmetries of the model. 
 
From the results of this paper we can
foresee the folllowing possible generalizations
and/or developments of.  
For example there is
an integrable hierarchy corresponding to an appropriate subset
$U \subset {\bf Q}$ which is different
from the set $U_{N}$ used in this paper.
 By ``appropriate" we mean that the
ideal of the ring ${\bf B}$ consisting of the operators annihilating
the set $h_{\gb} (t) ,\gb \in U$ is finitely generated. An example
of such a set is served by $\{ \gb \in {\bf Q} | deg (\gb ) \leq n 
\}$.

Another way to generalize our construction is to start with an
arbitrary finitely generated ${\bf Z}$-graded abelian group ${\bf G}$ and
fix a semigroup $\Q \subset \G$
in such a way that the ${\bf Z}$-gradation of $\G$ induces
${\bf Z}_{+}$-gradation of $\Q$.
Let us take for example
${\bf G} = {\bf Z} \times {\bf Z} \times {\bf Z} $, 
$\Q= \{ (m,n,p) \in \G \mid m ,n,p \geq 0 \}$.
 In the simplest case analogous
to the case $N=1$ considered in the paper we will have
three wave operators $W_{i} = \partial _{i} + a_{i} (t), i=1,2,3$ 
subject to constraints $[W_{i}, W_{j} ]=0$. These constraints
have a solution $a_{i} (t) = \partial _{i} \Psi (t)$ and
one of the simplest equations of corresponding hierarchy
of equations written in terms of the potential $\Psi$ reads
\begin{eqnarray*}
\frac{\partial{\Psi}}{\partial t_{(1,1,1)}} + 
(\partial _{x} \Psi ) (\partial _{y} \partial _{z} \Psi )+
(\partial _{y} \Psi ) (\partial _{x} \partial _{z} \Psi )+
(\partial _{z} \Psi ) (\partial _{x} \partial _{y} \Psi ) \\
= (\partial _{x} \partial _{y} \partial _{z} \Psi ) + 
(\partial_{x} \Psi )(\partial_{x} \Psi )(\partial_{x} \Psi ).
\end{eqnarray*}
This equation has the following solution:
\begin{eqnarray*}
\Psi = ln \left(\int du d v d w ~~\Phi (u,v,w) e^{xu +yv +zw
+uvwt_{(1,1,1)}}\right),
\end{eqnarray*}
which can be verified by direct substitution.

It is worth mentioning that integrable
systems in $2+1$ dimensions
which are different from KP hierarchy have been investigated
recently using Lax formalism (see e. g. \cite{Ablowitz}
and \cite{Zakharov} ): 
generalized Schroedinger equations,   
Davey- Stewartson equation 
(a generalization of non-linear 
Schroedinger equation to (2+1)
dimensions) and their
close relatives. It will be interesting
to study the Lax structure of $2+1$-dimensional
Burgers hierarchy ( extended-$KP(1)$) 
introduced in Section 6.

The following question seems to be extremely interesting
to explore: is there
a possible connection between NMM and $multidimensional$
gravity analogous to the well-known relations between
HMM and two-dimensional gravity? To address this question
we need to formulate an analog of
Feynman rules for the computation of
the coefficients of the asymptotic expansion of
the partition function of NMM  in the case when coupling
constants are large. Corresponding analysis is complicated
by the non-linearity of the space of normal matrices, but
there are no principle obstructions for the progress.  

The asymptotic expansion of the partition function
of the hermitian matrix model (with a non-polynomial
potential) carries topological information
about the moduli space of Riemann surfaces \cite{Penner}. 
What can be said in this respect about
the partition function of NMM?
 
As far as algebraic geometry is concerned it will
be interesting to study an analog
of quasiperiodic solutions to KP hierarchy \cite{Krichever}
in the context of extended-$KP(N)$ hierarchies. Such
solutions could correspond to algebraic surfaces on the one
hand and to points of $Gr(N, \infty)$ on the other,
thus providing us with a sort of $multidimensional$
Krichever construction,  \cite{Schwarz}.

\section{Appendix}

\subsection{The proof of Lemma 1}
Recall that wave operator $W_{\ppb}$ defined by (\ref{91}) is of order 
$N$
in $\partial _{x}$ while $W_{\qb}$ from  
 (\ref{92})  is of the first order in $\partial _{y}$.
Any operator $O \in {\bf B}$ can be presented in the form
\begin{eqnarray*}
O= \sum_{m_{1}=0}^{M_{1}} \sum_{m_{2}=0}^{M_{2}} C_{m_{1},m_{2}}
\partial_{x} ^{m_{1}} \partial _{y} ^{m_{2}},
\end{eqnarray*} 
where coefficients $C$'s are elements of ${\bf C} [[t]]$.
Therefore, 
\begin{eqnarray*}
O = \sum_{m_{1}}^{M_{1}} C_{m_{1}, M_{2}}
\partial_{x} ^{m_{1}} \partial _{y} ^{M_{2}} + \mbox{terms
of lower order in } \partial _{y}. 
\end{eqnarray*}
Or, 
\begin{eqnarray*}
O = \sum_{m_{1}}^{M_{1}} C_{m_{1}, M_{2}}
\partial_{x} ^{m_{1}} \partial _{y} ^{M_{2}-1} W_{\qb} + \mbox{terms
of lower order in } \partial _{y}.
\end{eqnarray*}
So, applying induction we see that
\begin{eqnarray*}
O  = b_{\qb} W_{\qb} + O' ,
\end{eqnarray*}
where $O' \in \B$ is of zeroth order in $\partial _{y}$. 
If the degree of $O'$ with respect to $\partial _{x}$ is less
than $N$ we are done. If not, we have $O'=
C_{K} \partial _{x}^{K}+$ lower order terms 
$=C_{K} \partial _{x} ^{K-N} W_{\ppb}+$
lower order terms. If the order of the 
omitted terms is greater than $N$ we continue
the process. After a finite number of
steps we will arrive at the following 
representation of the operator $O'$:
\begin{eqnarray*}
O' = b_{\ppb} W_{\ppb} +\sum_{\gb \in U_{N}}
c_{\gb} (t) \dv ^{~\gb},
\end{eqnarray*}
where $b_{\ppb}$  is a differential operators in $\partial _{x}$
only. Therefore an operator $O$ can be presented as follows:
\begin{eqnarray}
O  = b_{\ppb} W_{\ppb} + b_{\qb} W_{\qb} + \sum_{\gb \in U_{N}}
c_{\gb} (t) \dv ^{~\gb }.
\label{basic}
\end{eqnarray}
Applying equality (\ref{basic} ) to the function $h_{\hb} (t)$
with $\hb \in U_{N}$ we see that 
\begin{eqnarray}
\sum_{\gb \in U_{N}}c_{\gb} (t) (\dv ^{\gb } h_{\hb}) =0 ,~ \hb \in U_{N} .
\label{S}
\end{eqnarray}
But (\ref{S} ) is a system of linear algebraic homogeneous
equations with respect to
$\{ c_{\gb } \} _{\gb \in U_{N}}$  with the determinant which is an
invertible formal power series.
Thus $c_{\gb} =0 , \gb \in U_{N}$
and Lemma 1 is proved. 

\subsection{The proof of Lemma 2 }

First of all let us introduce the grading
of the ring ${\bf B}$ by
setting $deg (\partial _{x}) =1$, $deg (\partial _{y} ) =N$.
The degree of an element of ${\bf B}$ is then by definition the maximal
degree of monomials in the linear combination constituting this element.
We will prove Lemma 2 basing on the following proposition:
\begin{lm}
Any element $O \in \B$ admits a unique representation
of the form
\begin{eqnarray}
O = \sum_{n_{1}, n_{2} \geq 0} c_{n_{1}, n_{2}} 
W_{\qb}^{n_{2}} W_{\ppb}^{n_{1}},
\label{un}
\end{eqnarray}
where $deg(c_{n_{1}, n_{2}}) <N$ and $c_{n_{1}, n_{2}} = 0$
if $n_{1}  >> 0$ or $n_{2} >> 0$.   
\end{lm}
We will prove Lemma 4 later. Now let us show how the statement of Lemma 2
can be deduced from Lemma 4. 

Consider the following sequence of maps:
\begin{eqnarray}
0 \rightarrow \B \stackrel{\alpha}{\rightarrow}
 \B \oplus \B \stackrel{\beta}{\rightarrow} \I \rightarrow 0 ,
\label{exseq}
\end{eqnarray}
where $\alpha (b) = \biggl(-b (W_{\qb} + Y) , b W_{\ppb} \biggr),
 b \in \B$
and $\beta (c_{1} , c_{2} ) = c_{1} W_{\ppb} + c_{2} W_{\qb},
(c_{1} ,
c_{2} ) \in 
\B \oplus \B$. 
Lemma 2 is equivalent actually to the statement of exactness of this
sequence.  So, let us check the exactness of (\ref{exseq}).

First, we see that the map $\alpha$ is monomorphism. Really, $\alpha (b) 
=0$ implies
in particular that $b W_{\ppb} =0 $ which yields $b=0$, as
the ring $\B$ is an integral domain. Thus, $Ker (\alpha) =0$.
Second, the map $\beta$ is epimorphism. Really, it follows from Lemma 1
that for any $O \in \I$ there are $c_{1}, c_{2} \in \B$ such that 
$O= c_{1} W_{\ppb} + c_{2} W_{\qb} = \beta (c_{1} , c_{2})$.
Thus, $\beta$ is onto.

Finally, we have to verify that $Im (\alpha ) = Ker (\beta )$. It is easy
to see that $Im (\alpha ) \subset Ker (\beta )$. Really, $\beta \cdot
\alpha (b)
=\beta \biggl(( -b \cdot (W_{\qb} +Y ) , b \cdot W_{\ppb} )\biggr) 
= b \cdot ( [W_{\ppb} , W_{\qb} ]
- Y W_{\ppb} )$, which is zero in virtue of the relation (\ref{114}).
Let us prove the opposite inclusion, $Ker (\beta ) \subset Im (\alpha ) 
$.
If  $(c_{1} , c_{2} ) \in Ker (\beta )$ then for any $b \in \B$
$(c_{1} , c_{2} ) - \alpha (b) = 
\biggl(c_{1} + b (W_{\qb} +Y) , c_{2} - b W_{\ppb} \biggr) \in Ker 
(\beta)$.
By Lemma 4, $c_{2} = \sum_{n_{1}, n_{2} \geq 0} ~ q_{n_{1} , n_{2}} 
W_{\qb} ^{n_{2}} W_{\ppb} ^{n_{1}}$. Choose
$b =  \sum_{n_{1}, n_{2} \geq 0} ~ q_{n_{1}+1 , n_{2}} 
W_{\qb} ^{n_{2}} W_{\ppb} ^{n_{1}}$.   Then,
$(c_{1} , c_{2} ) - \alpha (b) = \biggl(c_{1} + b (W_{\qb} +Y ), 
\sum _{n \geq
0}
~ q_{0, n} W_{\qb} ^{n} \biggr) \in Ker (\beta )$. Again, by Lemma 4,
$c_{1} + b (W_{\qb} + Y ) = \sum_{n_{1}, n_{2} \geq 0} ~ p_{n_{1} , n_{2}} 
W_{\qb} ^{n_{2}} W_{\ppb} ^{n_{1}}$. So,
 \begin{eqnarray*}
\sum_{n_{1}, n_{2} \geq 0} p_{n_{1} , n_{2}} 
W_{\qb} ^{n_{2}} W_{\ppb} ^{n_{1} +1} +
\sum _{n \geq 0}q_{0, n} W_{\qb} ^{n+1}=0
\end{eqnarray*}
But the l.h.s. of the equation above is a representation of $0$
in the form (\ref{un}), it follows then from the uniqueness of such 
representation
that $p_{n_{1} , n_{2}} =0$ for $n_{1} , n_{2} \geq 0$ and $q_{0,n} =0$
for $n \geq 0$.
Going back we conclude that $(c_{1} , c_{2} ) - \alpha (b)=0$, which
proves that $Ker (\beta) \in Im (\alpha)$.

It remains to prove Lemma 4 which states the existence
and uniqueness of decomposition (\ref{un}).
Take any element $O \in \B$ of
order less than $(n+1) \cdot N$ with $n \geq 0$. It can be presented in the
following
form:
\begin{eqnarray}
O= \sum_{k=0}^{n} q_{k} \partial _{x} ^{k \cdot N} \partial _{y} ^{n-k}
+O',
\label{159}
\end{eqnarray}
where the degree of $q_{k}$ is less than $N$ and the degree
of $O'$ is less than $k \cdot N$. Thus,
\begin{eqnarray}
O- O'=q_{0} W_{\qb}^{n} +q_{0} (\partial _{y} ^{n}-W_{\qb}^{n})
 + \cdots +q_{n} \partial _{x} ^{n \cdot N}  \nonumber\\
= q_{0} W_{\qb} ^{n} + q'_{1}  \partial _{y} ^{n-1}\partial _{x} ^{N}  
+\cdots
+q'_{n} \partial _{x} ^{n \cdot N} + D_{1},
\label{base}
\end{eqnarray}
where $deg(q' _{k}) < N$ and $deg( D_{1}) < n \cdot N$.

Suppose that we have proved that for some $m$ such
that $ 0 \leq m < n,$
\begin{eqnarray}
O- O'=c_{0} W_{\qb} ^{n} + \cdots + c_{m}  W_{\qb} ^{n-m}
W_{\ppb} ^{m}  + d_{m+1} \partial_{y} ^{n-m-1} \partial _{x} ^{(m+1) 
\cdot
N} +
\cdots +d_{n} \partial _{x} ^{n \cdot N} + D_{m+1},
\label{ind1}
\end{eqnarray}
where degrees of operators $c_{i}$ and $d_{j}$ are less than $N$ and
$deg(D_{m+1}) < N$. Then we see that 
\begin{eqnarray*}
O- O'&=&c_{0} W_{\qb} ^{n} + \cdots + c_{m}  W_{\qb} ^{n-m}
W_{\ppb} ^{m \cdot N}  + c_{m+1} W_{\qb}^{n-m-1} W_{\ppb} ^{(m+1) \cdot N}
\\
 &+& d_{m+1} (\partial_{y} ^{n-m-1} \partial _{x} ^{(m+1) \cdot N} -
 W_{\qb}^{n-m-1} W_{\ppb} ^{(m+1) \cdot N)})+
\cdots +d_{n} \partial _{x} ^{n \cdot N} + D_{m+1} \\
 &=&c_{0} W_{\qb} ^{n} + \cdots + c_{m+1}  W_{\qb} ^{n-m-1}
W_{\ppb} ^{(m+1) \cdot N}  + d'_{m+2} \partial_{y} ^{n-m-2} \partial _{x} ^{(m+2)
\cdot N} \\
&+& \cdots +d'_{n} \partial _{x} ^{n \cdot N} + D_{m+2},
\end{eqnarray*}
where degrees of operators $c_{i}$ and $d'_{j}$ are less than $N$ 
and $deg(D_{m+1}) < N$. At some point we defined $c_{m+1} = d_{m+1}$.
Thus we proved by induction, the base of which is (\ref{base}) and the 
induction
hypothesis is (\ref{ind1}), that
\begin{eqnarray}
O= \sum_{k=0}^{n} c_{k} W_{\qb} ^{k} W_{\ppb} ^{n-k}+O'',
\label{ind2}
\end{eqnarray}
where $deg(O'') < n \cdot N$. Note that if $deg(O) < N$, then
the existence of (\ref{un}) is clear. So, using (\ref{ind2}) to
generate induction in degree we verify the existence
of representation (\ref{un}) for any $O \in \B$.

To prove the uniqueness of (\ref{un}) we must show that
the equality 
\begin{eqnarray}
\sum_{n_{1}, n_{2} \geq 0} c_{n_{1} , n_{2}} 
W_{\qb} ^{n_{2}} W_{\ppb} ^{n_{1}}=0
\label{zero}
\end{eqnarray}
 implies $c_{n_{1} , n_{2}} =0$
for all $ n_{1}, n_{2} \geq 0$. We assume of course
that $deg(c_{n_{1} ,n_{2}}) < N$.
Suppose that $n \cdot N \leq deg( \sum_{n_{1}, n_{2} \geq 0} c_{n_{1} ,
n_{2}} 
W_{\qb} ^{n_{2}} W_{\ppb} ^{n_{1}}) < (n+1) \cdot N$ with $n > 0$. 
Then
\begin{eqnarray*}
W_{\qb} ^{n_{2}} W_{\ppb} ^{n_{1}} \equiv
\sum_{n_{1}+ n_{2}=n} c_{n_{1} , n_{2}} 
W_{\qb} ^{n_{2}} W_{\ppb} ^{n_{1}} + O''' =0,
\mbox{ with } deg(O''') < n \cdot N.
\end{eqnarray*}
Thus $c_{0, n} \partial _{y} +($ terms of lesser degree in $\partial _{y} 
) =0$. Therefore, $c_{0,n} =0$. Continuing step-by-step considerations of
terms of the highest  degree we see that $c_{n_{1}, n_{2}}=0$ for
$n_{1} +n_{2} =n$. But this contradicts the assumption that
$deg( \sum_{n_{1}, n_{2} \geq 0} c_{n_{1} , n_{2}} 
W_{\qb} ^{n_{2}} W_{\ppb} ^{n_{1}}) \geq n \cdot N$
with $n >0$. Thus all coefficients $c_{n_{1}, n_{2}}$  in the l.h.s.
of (\ref{zero}) for
$n_{1} +n_{2} >0$ are equal to $0$ and we are left with
the equality $c_{0,0} =0$. The uniqueness of (\ref{un})
is thus proved and so is Lemma 4. 
 
\subsection{The proof of Lemma 3}
First we will construct $N$ linearly independent elements
of  $\bf{C} [[t]]$ annihilated by $W_{\ppb}$ and $W_{\qb}$ and
satisfying condition (\ref{133x}).
Then we will prove that any element of $\bf{C} [[t]]$ annihilated by
$W_{\ppb}$ and $W_{\qb}$ is a linear combination of these elements
with coefficients independent of $x,y$.
  
Let $\tilde{h}_{1}, \cdots , \tilde{h}_{N}$ be $N$ linearly independent 
elements
of $\bf{C} [[t]]$ generating $KerW_{\ppb}$.
They can be chosen to have the following form: $\tilde{h}_{k} =
x^{k-1}+ (\mbox{ higher order terms in } x)$ 
with $k=1, \cdots , N$, so that the condition
(\ref{133x}) is satisfied.  
Applying  (\ref{114})
to these elements,
we see that $W_{\qb}\tilde{h}_{i} (t) \in 
KerW_{\ppb}$. 
Therefore, there exists an $N \times N$ matrix $E (t)$ independent from 
$x$
such that 
\begin{eqnarray}
W_{\qb} \tilde{h}_{i} (t) = \sum_{j} E(t)^{i}_{j} \tilde{h}_{j} (t)
\label{ann}
\end{eqnarray}
Consider an invertible matrix $F(t)$ solving the equation
\begin{eqnarray}
\biggl( \partial _{y} F(t) \biggr) \cdot F(t)^{-1} = E(t).
\label{hol}
\end{eqnarray}
Such a solution always exists in the formal category: it is
easy to verify that $F(t)= I +F_{1}(t)\cdot y
 +F_{2}(t) \cdot y^{2} + \cdots$, where $I$ is an $N \times N$
identity matrix and $F_{m} (t), m > 0$ are $N \times N$
matricies independent of $y$. They are determined
one-by-one from the recursion relation
$F_{m+1} (t) = \frac{1}{m+1} \sum_{i+j=m} E_{i} (t) \cdot F_{j} (t)$ with
$m \geq 0$,  $E(t) \equiv \sum_{i \geq 0} E_{i} (t) y^{i}$ and
$F_{0} (t) \equiv I$. The evaluation of our solution on complex
numbers for $y$ can be written as usual in the form of path-ordered 
exponent.

Consider now a set of functions $h_{1}, \cdots , h_{N}$ defined
from: $\tilde{h}_{i}(t) = \sum_{j} [F(t)] _{i}^{j}  h_{j} (t),~ i=1, 
\cdots , N$.
Then  $W_{\qb} h_{i } (t) =0,~ i=1, 
\cdots , N$ due to
(\ref{ann}) and the fact that the matrix
$F(t)$ solves equation (\ref{hol});
and $W_{\ppb} h_{i } (t) =0, i=1, \cdots , N$ due to the fact that
$F(t)$ is independent
from $x$. Thus  $h_{1}, \cdots , h_{N} \in KerW_{\ppb} \bigcap 
KerW_{\qb}$. Due to the non-degeneracy of the matrix
$F(t)$ and the fact that the elements
$\tilde{h}_{1}, \cdots , \tilde{h}_{N}$ 
were chosen to satisfy 
the condition (\ref{133x})
it follows that elements  $h_{1}, \cdots , h_{N}$
also satisfy the condition (\ref{133x}) 

Moreover, these elements generate the intersection of the kernels: 
suppose
there is $f(t) \in \bf{C} [[t]] $ such that $f(t) \in KerW_{\ppb} \bigcap 
KerW_{\qb}$.
Then $f(t) \in KerW_{\ppb} $ as well. Therefore, there are coefficients 
$d_{1} (t),
\cdots d_{N} (t)$ independent of $x$ such that 
\begin{eqnarray}
f(t) = \sum_{i} d_{i} (t) h_{i} (t).
\label{dec}
\end{eqnarray}
Thus all we have to prove is that coefficients $d_{i}$'s 
are independent from $y$. Applying $W_{\qb}$
to (\ref{dec}) we obtain that $\sum_{i} \partial _{y} \biggl(d_{i} (t) \biggr)
h_{i} (t) = 0$. Then by linear independence of $h_{i}$'s
over formal power series depending on $y$ and $t_{\gb}, \gb \in 
\Q \setminus \ppb , \qb$ we conclude $\partial _{y} (d_{i} (t)) = 0 , i= 1, \cdots, 
N$.
Lemma 3 is proved.

\subsection{The proof of bosonization relation } 

Here we present the proof of (\ref{eqn;ninetynine}).
First, consider operators $P_{+} (-t) \equiv e^{-H(t)} P_{+} e ^{H(t)}$
and $P_{-} (-t) \equiv e^{-H(t)} P_{-} e^{H (t)}$.  Following \cite{JM1},
we will call them {\em Clifford operators}.  The identities 
below \cite{JM2} express the properties of the Clifford operators 
which will be important for 
us:
\begin{eqnarray}
\lv P_{+} (-t) = \lv e^{H(t)} =\lv P_{-} (-t).\label{eqn;seventy}
\end{eqnarray}

The proof of (\ref{eqn;seventy}) is not completely straightforward
and was not explained in \cite{JM2}. Therefore let us outline
the proof.

Consider the state
\begin{eqnarray}
\langle \epsilon | = \lv e^{H (\epsilon t)} P_{+}
(-t).\label{eqn;seventya}
\end{eqnarray}
We wish to compute $\frac{d}{d \epsilon } \langle \epsilon |$. 
Note that $H(t) \equiv e^{-H (\epsilon t )} H(t) e ^{H (\epsilon t ) }=
\sum_{\gb  \in \Gp} t_{\gb } J_{\gb } (- \epsilon t)$,
where we used the definition (\ref{eqn;fortyone})
and the agreement (\ref{eqn;fortyfour}).
Therefore
\begin{eqnarray*}
\frac{d}{d \epsilon}  \lv e^{H (\epsilon t)} H(t)  P_{+} (-t) & & \\
&=& \sum_{\gb  \in \Gp} t_{\gb } 
\sum_{\hb \in \Gp} \lv e^{H (\epsilon t)} \f _{\hb } (-\epsilon t) 
\fb _{\gb  +\hb } (-\epsilon t) P_{+} (-t)  \\
&=& \sum_{\gb  \in \Gp } t_{\gb } \sum _{\hb  \in \Gp} \lv \f _{\hb } \fb 
_{\gb 
+\hb }
e^{H (\epsilon t )} P_{+} (-t) =0,
\end{eqnarray*}
where the first equality is due to the properties
(\ref{eqn;fifty})-(\ref{eqn;fiftyone}) of projector
operators and the last equality follows from (\ref{eqn;thirtysix}). 
Therefore
the
state $\langle \epsilon | $ is in fact independent of $\epsilon$. 
Finally,  observing that $\lv P_{+} = \lv$ we obtain the desired result:
\begin{eqnarray*}
\lv e^{H (t) } \;=\; \lv e^{H(t)} P_{+} (-t) \;=\; \langle \epsilon =1 
|
\;=\;
\langle \epsilon =0 | \;=\; \lv P_{+} (-t).\label{eqn;seventyc}
\end{eqnarray*}
The first equality in (\ref{eqn;seventy}) is proved.  
The second one can be verified along the same
lines.  

Next we need to prove that
\begin{eqnarray}
\llun e^{H(t)} \f (\zv \,) A \Pp =  \X 
\zv ^{~(N-1) \cdot \ppb  } 
 \llun  \f _{(N-1) \cdot \ppb} e^{H (t)} A \Pp . 
\label{eqn;seventythree}
\end{eqnarray}
The proof is the following. 
Using (\ref{evol}) we can perform
the following computation: 
\begin{eqnarray*}
[ \Pp (-t) \Uh (-t),  \f (\zv \,) ] & A \Pp & \\ 
&=& e ^{V ( t,  \zv \, ) }
e^{-H ( t )} [ \Pp \Uh,  \f (\zv\,) ] 
 e^{H (t )} A\Pp  \\
&=& e^{V (t,  \zv\,)}
e^{-H (t) } 
\zv ^{~(N-1) \cdot \ppb } \sum_{n \geq 0}
\zv ^{\, -n \cdot \ppb}
[\Pp \Uh , \f _{(N-1- n) \cdot \ppb}]
e^{H (t)} A \Pp  \\
&=& e^{V (t,  \zv\,)}
e^{-H (t-\mu) } 
\zv ^{~(N-1) \cdot \ppb } 
[\Pp (-\mu ) \Uh ( -\mu ) ,\f _{(N-1) \cdot \ppb}]
e^{H (t- \mu )} A \Pp  \\
&=& \X  e^{-H (t) } 
\zv ^{~(N-1) \cdot \ppb } 
[\Pp (-\mu ) \Uh ( -\mu ) ,\f _{(N-1) \cdot \ppb}]
e^{H (t )} A \Pp, 
\end{eqnarray*}
where
$\prod_{\gb  \in U_{N}} \fb_{\gb } = \hat{U} _{N}$ and 
$\mu _{\gb} = \frac{1}{n}$ if $\gb = n \cdot \ppb$ with $n> 0$
and $\mu _{\gb} = 0$ otherwise. 
A calculation shows that
\begin{eqnarray*}
\Pp (- \mu )  =
\Pp \cdot e^{\left( \zv^{\,-\ppb } \sum_{\gb  \in {\bf G}_{-1}} \f _{\gb }  
\fb
_{\gb  +\ppb } (- \mu) \right) }
\end{eqnarray*}
Using this result and observing that $\f _{\hb } \Pp =0,  \hb  \in \Gm$
we find that
\begin{eqnarray}
[\Pp (-t ) \Uh (-t ) ,  \f (\zv \,) ] 
A \Pp = 
\X  \zv ^{~(N-1) \cdot \ppb  }  e^{-H(t)} \Pp
[ \Uh (-\mu ), \f _{(N-1) \cdot \ppb } ] e^{H (t)} A \Pp .
\label{eqn;seventyfive}
\end{eqnarray}
Applying the operator equality (\ref{eqn;seventyfive}) to the left vacuum 
and
using
the property (\ref{eqn;seventy}) of Clifford operators we see that
\begin{eqnarray}
\lv  \Uh e^{H(t)} \f (\zv ) A \Pp =
\X \zv ^{(N-1) \cdot \ppb} \lv \Uh \f _{(N-1) \cdot \ppb} e^{H(t)} A \Pp,
\label{last}
\end{eqnarray}
where we used that $\lv \Uh (-\mu ) \f _{(N-1) \cdot \ppb} =
\lv \Uh \f _{(N-1) \cdot \ppb}$, which can be verified directly.
Taking into account that $\lv \Uh \equiv \llun$, we obtain
(\ref{eqn;seventythree}).

Finally let us apply (\ref{eqn;seventythree}) to $\run$ and 
use the fact that $\Pp \run = \run$. Dividing
the result by $\tau (U_{N}, A, t)$ we arrive at the equation
(\ref{eqn;ninetynine}).  

\section{Acknowledgments } We are grateful
to D. Fuchs, J. Hunter, A. Konechny, 
G. Kuperberg, M.  Mulase, A.  Schwarz
and P. Vanhaecke
for numerous discussions and reading the manuscript. 
This research was partially supported by the U.  S. 
Department of Energy and National Science Foundation.


\begin{thebibliography}{Morosov}

\bibitem{Ablowitz} M. Ablowitz and A. Fokas,
{\em Method of solution for a class of
multidimensional non-linear evolution
equations,} Phys. Rev. Lett. {\bf 51} (1), p. 7 (1983);

\bibitem{BZ0} E. Brezin and A. Zee, 
{\em Universality of the correlations between eigenvalues
of large random matrices},
Nucl. Phys.  {\bf B402} 
p. 613 (1993);

\bibitem{BZ} E.  Brezin and A.  Zee,    {\em Universal relation between
Green's functions in random matrix theory}, e-print archive:
cond-mat/9507032,
Nucl. Phys.  {\bf B453}, p. 531 (1995);

\bibitem{Burgers} J. Burgers, {\em The non-linear diffusion
equation: asymptotic solutions and statistical problems},
Dordrecht-Holland; Boston: D. Reidel Pub. Co., 1974;

\bibitem{Chau 1} L. -L.  Chau and Y.  Yu,  
{\em Unitary polynomials in normal matrix model and
wave functions for the fractional quantum Hall effect},  
Phys.  Lett.   {\bf A167},  p.  452 (1992);

\bibitem{Chau 2} L. -L.  Chau and O.  Zaboronsky,  {\em
Normal matrix model,  Toda lattice hierarchy,  and the
two-dimensional electron gas in the strong magnetic field}, 
Proceedings in memory of professor Wolfgang Kroll, ed.
J. P. Hsu et. al.,
World Scientific, Singapore, 1997;

\bibitem{Cole} J. Cole, Quart. Appl. Math. {\bf 9} p. 225 (1951);

\bibitem{JM2} E. Date, M.  Kashiwara and T.  Miwa,  
{\em Transformation groups
for soliton equations. II}, Proc. Japan Acad. {\bf 57}, Ser. A,
p.387 (1981);

\bibitem{JM3} E. Date, M. Jimbo M.  Kashiwara and T.  Miwa,
{\em Transformation groups for soliton equations.  III},
J. Phys. Soc. Jpn., {\bf 50},  Ser.  A,  
p. 342 (1981);

\bibitem{JM} E.  Date,  M.  Jimbo,  M.  Kashiwara,  and T.  Miwa,  
{\em Integrable systems},  Proc.  RIMS Symp.  ``Non-linear
integrable systems - classical and quantum theory",  p.  39,  1983;

\bibitem{Dubrovin} B. Dubrovin, A. Fomenko, S. Novikov,
{\em Modern geometry - methods and applications,} New-York:
Springer-Verlag, 1990; 

\bibitem{Dyson} F. Dyson, {\em Statistical theory
of energy levels of complex systems}, I, II and III,
Jour. Math. Phys. {\bf 3}, p. 140, p. 157 and p. 166 (1962);

\bibitem{Fuchs} A. Fomenko, D. Fuchs, and V. Gutenmacher, 
{\em Homotopic topology,}   
Budapest : Akademia Kiado : [Distributors,
Kultura], 1986;

\bibitem{Schwarz} L. Friedlander and A. Schwarz, 
{\em Grassmannian and elliptic operators}, 
e-print archive: funct-an/9704003;

\bibitem{Ambjorn}  M. Fukuma, H. Kawai and R. Nakayama,
{\em Continuum Schwinger-Dyson equations and
universal structures in two-dimensional quantum
gravity,} Int. J. Mod. Phys. {\bf A6}, p. 1385 (1991);

\bibitem{Gasper} G. Gasper, M. Rahman,
{\em Basic hypergeometric series,}    Cambridge
[England] ; New York : Cambridge University Press, 1990,

\bibitem{Gurb} S. Gurbatov, A. Malakhov, A. Saichev, {\em
Nonlinear random waves and turbulence in nondispersive
media:waves, rays, particles,} New-York: Manchester University press, 1991;

\bibitem{Hopf} E. Hopf, Comm. Pure Appl. Math. {\bf 3} p. 201 (1950);

\bibitem{Itoyama} H. Itoyama and Y. Matsuo, 
{\em Noncritical Virasoro algebra of the $d < 1$ matrix
model and the quantized string field}, Phys. Lett. {\bf B255} p. 202 (1991); 


\bibitem{JM1} M.  Kashiwara and T.  Miwa,  {\em Transformation groups
for soliton equations.  I},  Proc.  Japan Acad. ,  {\bf 57},  Ser.  A,  
p. 342,  1981;

\bibitem{Kazakov} V. Kazakov, {\em The appearance of matter
fields from quantum fluctuations of $2D$-gravity},
Mod. Phys. Lett.  {\bf A4}, p. 2125 (1989); 

\bibitem{MMM} S.  Kharchev,   A.  Marshakov,   A.  Mironov and A.  
Morozov,  
{\em Generalized Kontsevich model versus Toda hierarchy and
discrete matrix models},   Nucl.  Phys.  {\bf B397} p.  339,   1993;
\bibitem{Mak2} Yu. Makenko, {\em Loop equations in
matrix models and in 2D gravity,} Mod. Phys. Lett. {\bf A6}, (no. 21),
p.1901 (1991);

\bibitem{Semenoff1} I. Kogan, G. Semenoff, 
{\em Fractional spin, magnetic moment and 
the short range interactions of anyons},
Nucl. Phys. {\bf B368} p. 718 (1992); 

\bibitem{Kolm} A. Kolmogorov, S. Fomin,
{\em Introductory real analysis}, New-York: Dover
Publications Inc., 1975;

\bibitem{Krichever} I. Krichever, Russ. Math.
Surveys, {\bf 32} p. 185;

\bibitem{Gurb1} N. Kuznetsov and B. Rozhdestvensky, 
ZhVMMF {\bf 1} (2), p. 217 (1961); 

\bibitem{Landau} L. Landau and E. Lifshitz,
{\em Fluid Mechanics}, 
London: Pergamon Press, 1987;
 
\bibitem{Leznov} A. Leznov and M. Saveliev,
{\em Theory of group representations and
integration of non-linear systems $X_{a, z \zb}=\exp{(kx)}_{a}$ },
Physica {\bf 3D}, p. 62 (1981);

\bibitem{Mehta} M. Mehta, {\em Random matrices,}
San-Diego: Academic Press, 1991;

\bibitem{Mironov} A.  Mironov and A.  Morozov,  
{\em On the origin of Virasoro constraints in matrix models:
lagrangian approach},  Phys.  Lett.   {\bf B252} p.  47,  1990;

\bibitem{Mohling} F.  Mohling,    {\em Statistical 
mechanics: methods and applications,  }
Jamaica, Queens, N. Y.: Publishers Creative services Inc. ,  1982;

\bibitem{Morosov} A. Morozov,    {\em Integrability and matrix 
models}, e-print archive:
hep-th/9303139, Phys.  Usp.   {\bf 1} p.  1,   1994; 

\bibitem{Mulase} M. Mulase, Math. Sci. {\bf 228} (1982)
and private communication;

\bibitem{Ohta} Y.  Ohta,  J.  Satsuma,  D.  Takakashi and
T.  Tokihiro,  {\em An elementary introduction to Sato theory}, 
Progress of Theoretical Physics Supplement {\bf 94}, p. 210  (1988);

\bibitem{Paffuti} G. Paffuti and P. Rossi,
{\em A solution to Wilson's loop equation
in lattice $QCD_{2}$}, Phys. Lett {\bf B92} p. 321 (1980);

\bibitem{Penner} R. Penner, {\em Perturbative series and the moduli
space of Riemann surfaces,} Comm. Math. Phys. {\bf 113},
p. 229 (1987);

\bibitem{Polyakov} A. Polyakov, {\em The theory of turbulence in two
dimensions.} Nucl. Phys. {\bf B396} p. 367 (1993);

\bibitem{Polyakov1} A. Polyakov, 
{\em Turbulence without pressure},
PUPT-1546, Jun 1995. 13pp.,
e-print archive: hep-th/9506189; 

\bibitem{SJM} M. Sato, T. Miwa and M. Jimbo,
{\em Holonomic quantum fields},
Publ. RIMS, Kyoto Univ., {\bf 14}, p. 223 (1978);
{\bf 15}, pp. 201, 577, 871 (1979); {\bf 16}, p. 531 (1980);  

\bibitem{SJM1}M. Sato, M. Jimbo, T. Miwa and Y. Mori,
{\em Holonomic quantum fields: the 
unanticipated link between deformation theory of
differential equations and quantum fields},
RIMS-305 (1979); Lausanne Math. Phys., {\bf119} (1979);  

\bibitem{Sato} M. Sato, {\em Soliton equations as dynamical systems
on infinite dimensional Grassmann manifold}, RIMS Kokyuroku, {\bf 439},
p. 30 (1981);

\bibitem{Sato1} M. Sato and Y. Sato (Mori), {\em Nonlinear partial
differential equations in Applied Science}, ed. H. Fujita, P. Lax and
G. Strang, Kinokuniya/North-Holland, Tokyo, 1983, p. 259;

\bibitem{Semenoff} G. Semenoff, {\em Anyons and 
Chern-Simons theory: a review},
PRINT-91-0208 (BRITISH-COLUMBIA), Feb
1991. 32pp. Presented at Karpacz Winter 
School for Theoretical Physics, Karpacz,
Poland, Feb 18 - Mar 1, 1991. 

\bibitem{Altshuler} B. Simons, P. Lee and B. Altshuler,
{\em Matrix models, one-dimensional fermions, and
quantum chaos}, Phys. Rev. Lett., {\bf 72} (1) p. 64 (1994); 

\bibitem{Sinai} Ya. Sinai, {\em Two results concerning
asymptotic behavior of solutions of the Burgers equation
with force,} Jour. Stat. Phys., {\bf 64}, p.1 (1991).

\bibitem{Ueno} K. Ueno and K. Takasaki, {\em Toda
lattice hierarchy. I and II},
Proceedings of the Japan Academy, Ser. A, {\bf 59} (5),
p. 167 and {\em ibid.}, {\bf 59} (6), p. 215 (1983);

\bibitem{Zakharov} V. Zakharov and S. Manakov,
{\em Construction of the multidimensional integrable
systems and their solutions}, Funktz. Anal Prilozh.,
{\bf 19} (2), p. 11 (1985).

\end{thebibliography}
\end{document}